\documentclass[fleqn,usenatbib]{mnras}
\usepackage[T1]{fontenc}
\usepackage{ae,aecompl}
\usepackage{graphicx}
\usepackage{amsmath}
\usepackage{amssymb}

\usepackage{hyperref}
\hypersetup{
    colorlinks=true,     
    urlcolor=blue
}

\title[Sunspot positions in the Dalton Minimum]{Sunspot positions 
from observations by Flaugergues in the Dalton Minimum}

\author[E. Illarionov \& R. Arlt]{
Egor Illarionov,$^{1, 2}$\thanks{E-mail: egor.illarionov@math.msu.ru}
Rainer Arlt$^{3}$
\\
$^{1}$ Moscow State University, Moscow, Russia\\
$^{2}$ Institute of Continuous Media Mechanics, Perm, Russia\\
$^{3}$ Leibniz Institute for Astrophysics Potsdam, Germany
}

\date{Accepted XXX. Received YYY; in original form ZZZ}

\pubyear{2022}

\begin{document}
\label{firstpage}
\pagerange{\pageref{firstpage}--\pageref{lastpage}}
\maketitle

\begin{abstract}
French astronomer Honor\'e Flaugergues compiled astronomical
observations in a series of hand-written notebooks 
for 1782--1830, which are preserved at Paris Observatory. 
We reviewed these manuscripts and encoded the records that contain sunspot
measurements into a numerical table for further analysis.  
All measurements are timings and we found three types of measurements
allowing the reconstruction of heliographic coordinates.
In the first case, the Sun and sunspots cross vertical
and horizontal wires, in the second case, one vertical and two 
mirror-symmetric oblique wires, and in the third case, a rhombus-shaped set of wires. Additionally, timings of two solar eclipses
also provided a few sunspot coordinates. 
As a result, we present the time--latitude (butterfly) diagram of the reconstructed
sunspot coordinates, which covers the period of the Dalton Minimum
and confirms consistency with those of Derfflinger and Prantner.
We identify four solar cycles in this diagram and discuss the observed
peculiarities as well as the data reliability.
\end{abstract}




\section{Introduction}
The Sun is known to exhibit a cyclic magnetic activity,
but it is important to note that it is far from being strictly
periodic \citep[e.g.][]{hathaway2015,Cameron2019}. While the individual cycles all appear to be
slightly different, their ensemble is also subject to
long-term modulations. The most prominent change to the
cyclicity is the Maunder Minimum in the second half of
the 17th and the beginning of the 18th century, in which very few sunspots were seen.
The Maunder Minimum has been extensively studied in terms 
of sunspot numbers or active days \citep[e.g.][]{nagovitsyn2007,vaquero_ea2015,gao2017,Hayakawa2021,carrasco_ea2022}, regarding the availability and reliability of reports \citep[e.g.][]{Hoyt1996, usoskin_ea2015,Carrasco_ea2021},
and through proxies of solar activity \citep[e.g.][]{miyahara_ea2004,berggren_ea2009,Usoskin2021,asvestari_ea2017, Brehm2021}.

The Dalton Minimum near 1800 is less pronounced and had been less studied until recently.
Interestingly, the 
number of observations is smaller -- especially just before the
Dalton Minimum --, but some observers cover long periods and it 
is time to exploit these sources in full in order to obtain a
detailed picture of the solar activity in the Dalton Minimum.
While previous studies have looked at the sunspot number or the 
group sunspot number \citep[e.g.]{Usoskin2003, Nielsen, Svalgaard2016,
carrasco2021} as powerful indices of solar activity, the recent
advances in obtaining sunspot positions from the original drawings 
of sunspots in the solar disk, including an appropriate determination 
of the sunspot group number, are enhancing our knowledge enormously
\citep{hayakawa_ea2020, Hayakawa_2021}. Sunspot
positions from just before the Dalton Minimum are available from 
Johann Staudacher in Nuremberg \citep{arlt2009}, Peter Horrebow and 
his collaborators in Copenhagen \citep{karoff_ea2019}, and Hamilton 
and Gimingham at Armagh Observatory\citep{arlt2009armagh}. 
The present paper aims to add further positional data to 
these efforts.

Cycle~4 is interesting, since it started in 1784 and either lasted
at least until about 1798 or was followed by a very weak cycle 
hardly detectable in sunspot numbers. The latter option
should exhibit spots at higher latitudes in the course of cycle~4.
\citet{usoskin_ea2009} found that a few spots indeed appeared at
higher heliographic latitudes from the middle of cycle 4, based on 
data from Armagh and Staudacher. 

It is also worth mentioning that the Dalton Minimum was first noticed not through sunspot records
but through the auroral records of John Dalton and so named afterwards \citep[e.g.][]{Silverman2021}. The present study is providing new positional data for 1788--1830, based on astrometric measurements.

\section{Data set}

We investigated digital images of hand-written books of astronomical observations by Honor\'e Flaugergues located in the library of Paris Observatory (Paris site). Table~\ref{tab:data} contains direct references to the books. The data cover the period of 1782 to 1830, with the first sunspot observations being in 1788 and running in parallel to the ones by Derfflinger \citep{hayakawa_ea2020} and Prantner \citep{Hayakawa2021}. The data recorded in the last years overlap with the observations by Schwabe \citep{arlt_ea2013} and may be interesting for calibration. 

We reviewed the images manually to extract a numerical format of all records related to solar transit timings. In this process, we also took into account textual remarks on the observing procedure (e.g.\ orientation of the telescope).
The numerical records are provided in a GitHub repository\footnote{\href{https://github.com/observethesun/Flaugergues.git}{https://github.com/observethesun/Flaugergues}}.

\begin{table}
\caption{Links to the digitized books of astronomical observations by Honor\'e Flaugergues
and shelf references for the original books.}
\label{tab:data}
\centering
\begin{tabular}{lll}
\hline
Years &  Link, preceded by        & Shelf mark \\
      &  https://bibnum.obspm.fr/ &             \\
\hline
1782--1798 & ark:/11287/WDC9B & F1/1\\
1798--1808 & ark:/11287/1hnHF & F1/2\\
1808--1816 & ark:/11287/d6gPS & F1/3\\
1816--1828 & ark:/11287/13HDQ & F1/4\\
1828--1830 & ark:/11287/2KV00 & F1/5\\
\hline
\end{tabular}
\end{table}

\section{Reconstruction of coordinates}

\subsection{Time mapping to universal time}

For the convenience of further calculations, we convert the recorded local times to the universal time system (UTC). This requires the
position of the observer and we assume that the observer's coordinates, corresponding to Viviers, France, are  $44\degr29'$~N and $4\degr41'$~E. Flaugergues estimated the coordinates of the observatory as $44\degr29'04''$~N and $4\degr40'56''$~E (after correcting from Paris reference to Greenwich reference) in the book of 1782--1798 (image~000023). Flaugergues mentioned he had observed also in Aubenas -- apparently at the county council -- which is a few km away from Viviers. We consider an accuracy of one arc minute good enough for our purposes and the expected accuracy of the observations.

Now, given the observer's position, we compute the UTC time of the local noon. Then we apply the time difference between the recorded time and the local noon to the UTC time of the local noon. Thus we obtain a mapping from the recorded time to the UTC time. In this procedure, we take into account that some records refer to astronomical days which start at noon of the corresponding civil day. Due to this fact it is possible that the converted days are $\pm1$ day from the actual date. In some cases we were unable to resolve this uncertainty and for simplicity always use the recorded dates.

In should be noted that in certain periods Flaugergues used sidereal time instead of mean solar time. In particular, sidereal time was used from 1809 December~23 to 1816 September~4, from 1820 June~1 to 1823 February~6, and from 1825 February~21 to 1825 September~3. We take this into account when converting the recorded times to UTC. 

\subsection{Horizontal and vertical wires}
\label{sec:hv}
\begin{figure}
\includegraphics[width=\columnwidth]{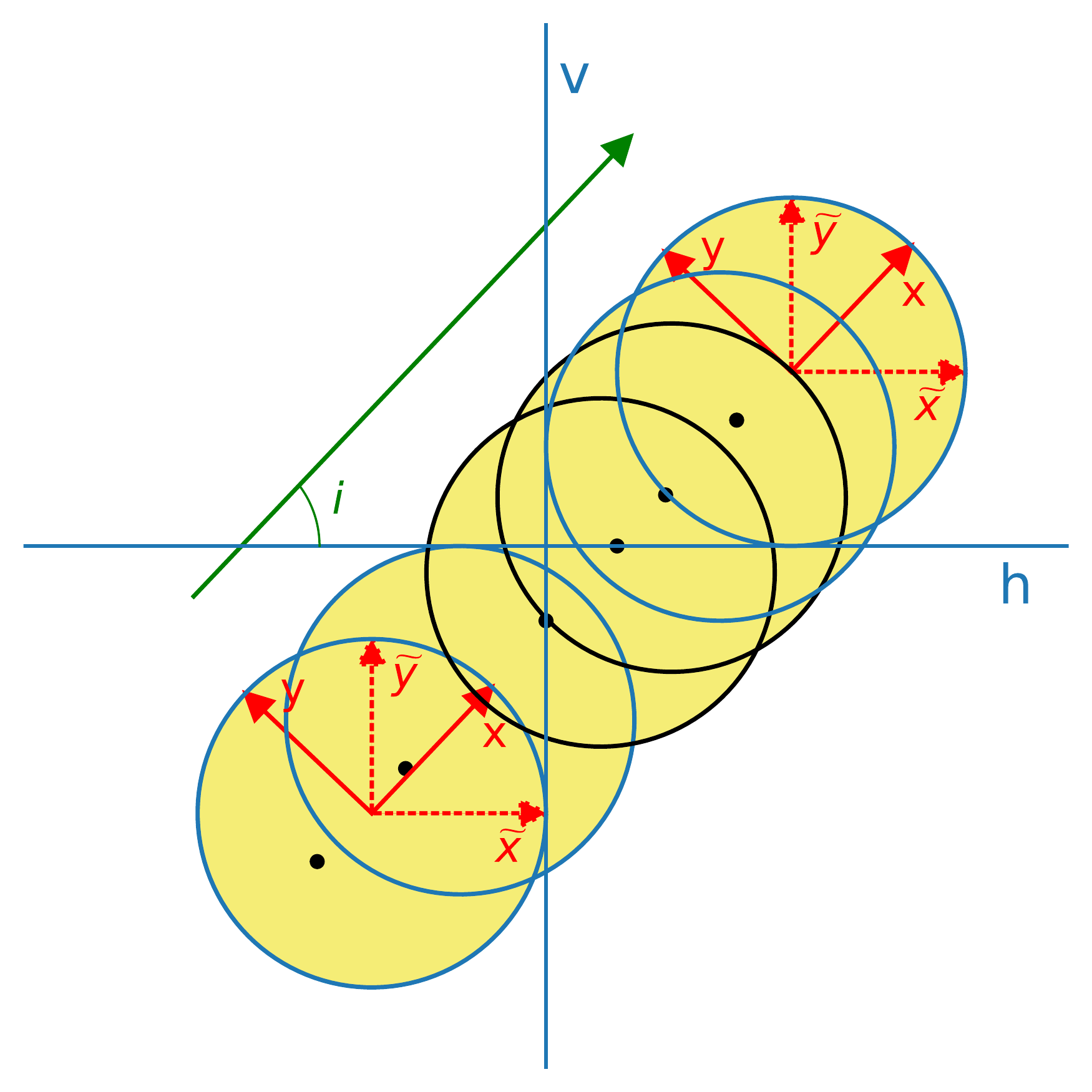}
    \caption{Sequence of transits of the solar disk and the sunspot (black dot) through the horizontal (h) and vertical (v) wires corresponding to Table~\ref{tab:hv_rec_1}. The blue circles show contacts of the solar disk with the wires. The black circles refer to contacts of the sunspot with the wires. The green arrow is parallel to the line of solar motion and indicates its direction. The inclination of this line towards the horizontal line is denoted as $i$. The red arrows show an $x,y$-coordinate system moving together with the solar disk. Note that the Sun is rising ($i < 0$).}
    \label{fig:fig1}
\end{figure}

Flaugergues used a classical cross-hair in the periods of 1796--1799
and 1824--1830. He called the wires horizontal and vertical since
indeed they are supposed to represent an azimuthal coordinate system.
Table~\ref{tab:hv_rec_1} shows an example of the transition times for the solar disk and the sunspot observed on 1796 July~7. Fig.~\ref{fig:fig1} illustrates these events. Note that we plot the events assuming the direct motion of the Sun, while Flaugergues could in fact have observed them upside down or mirrored.

\begin{table}
\caption{Observation on 1796 July~7 with passage times of two solar
limbs and a spot at a vertical and a horizontal wire.}
\centering
\begin{tabular}{llll}
\hline
Object & Event & Recorded time \\
\hline
     Sun &     v & 09:01:09 \\
     Sun &     h & 09:02:00 \\
     Spot &     v & 09:03:21.5 \\
     Spot &     h & 09:04:02.5 \\
     Sun &     v & 09:04:30 \\
     Sun &     h & 09:05:11 \\
     \hline
\end{tabular}
\label{tab:hv_rec_1}
\end{table}

To derive the heliographic coordinates of the sunspot, we first calculate the time differences for the solar disk transits:
\begin{eqnarray}
    \Delta_{\text{v}} &=& t(\odot\text{ at v second time}) - t(\odot\text{ at v first time}) \nonumber\\
    \Delta_{\text{h}} &=& t(\odot\text{ at h second time}) - t(\odot\text{ at h first time})
\label{eq:hv_deltas}
\end{eqnarray}

From equation~\ref{eq:hv_deltas}, we obtain an inclination angle (denoted as $i$) between the horizontal wire and the line of the solar motion: 
\begin{equation}
    |i| = \arctan(\Delta_{\text{v}} / \Delta_{\text{h}})
\label{eq:hv_alpha}
\end{equation}
Note that we assume that the angle $i$ is negative for the rising Sun and positive for the descending motion.

In particular, equation~\ref{eq:hv_deltas} and~\ref{eq:hv_alpha} allow the estimation of the solar ``radius'', measured in time units, by
\begin{equation}
    2\rho_\odot = \Delta_{\text{h}} \sin{|i|} = \Delta_{\text{v}} \cos{|i|}
\label{eq:hv_r}
\end{equation}
Note that this radius is still a time difference; the conversion to
or from an angular radius involves the declination of the Sun at 
a given moment.

Then we calculate the time differences between the sunspot and solar disk transits:
\begin{eqnarray}
    \delta_{\text{v}} &=& t(\text{Sunspot at v}) - t(\odot\text{ at v first time}) \nonumber\\
    \delta_{\text{h}} &=& t(\text{Sunspot at h}) - t(\odot\text{ at h first time})
\label{eq:hv_spot}
\end{eqnarray}

From equations~\ref{eq:hv_deltas} and~\ref{eq:hv_spot} we obtain $\widetilde{x}$- and $\widetilde{y}$-coordinates of the sunspot, where the $\widetilde{x}$-axis is parallel to the horizontal wire, the $\widetilde{y}$-axis is parallel to the vertical wire, and the radius of the Sun is unity:
\begin{eqnarray}
    \widetilde{x} &=& 1 - 2\delta_{\text{v}} /\Delta_{\text{v}}  \nonumber\\
    \widetilde{y} &=& 1 - 2\delta_{\text{h}} / \Delta_{\text{h}} 
\label{eq:hv_xy}
\end{eqnarray}
Note that the above expressions hold only if the Sun is rising.
Then we rotate the $(\widetilde{x}, \widetilde{y})$ coordinates by the inclination angle~$i$ 
to obtain the $(x, y)$ coordinates, where the $x$-axis is parallel to the motion of the Sun,
the $y$-axis points to the northern hemisphere, and radius of the Sun is unity. The latter coordinate system will
also be used for other types of measurements.
In particular, for the transits shown in Table~\ref{tab:hv_rec_1}, we obtain $\rho_\odot=69.23$~s, $i=-46.46\degr$,
$x=-0.42$, and $y=0.04$.

The final step is to apply the solar parameters $B_0$, $L_0$, and $P$, computed for the time of observation, in order to obtain the latitude $b$ and the Carrington longitude $L$. In particular, for the sunspot in Table~\ref{tab:hv_rec_1}, we obtain $b=5.9\degr$, $L=285.6\degr$.

\begin{figure}
\includegraphics[width=\columnwidth]{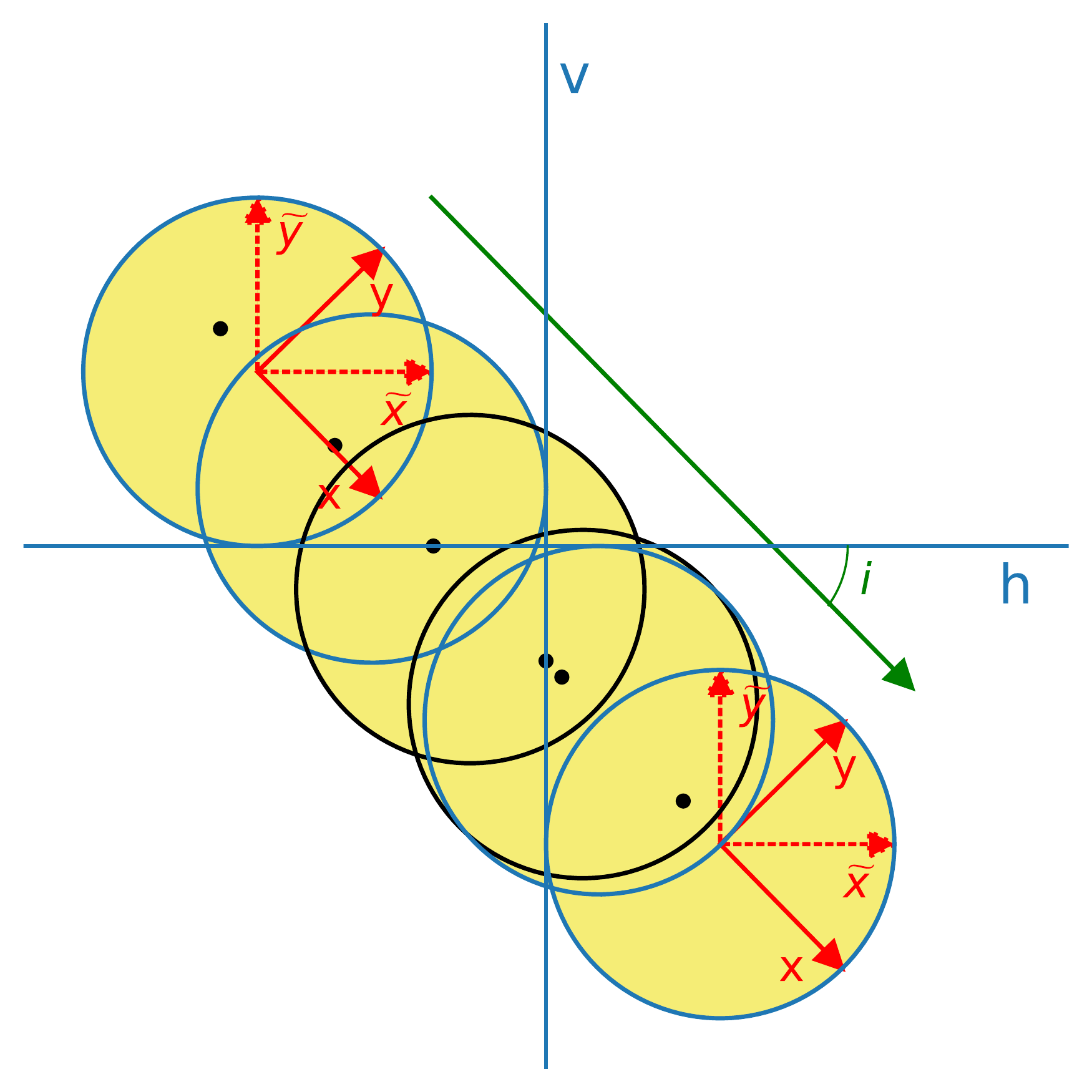}
    \caption{Sequence of transits of the solar disk and the sunspot corresponding to Table~\ref{tab:hv_rec_2}. The notation is the same as in Fig.~\ref{fig:fig1}. Note that the Sun is setting ($i > 0$).}
    \label{fig:fig2}
\end{figure}

Now we discuss the modifications for the setting Sun. Table~\ref{tab:hv_rec_2} shows an example of a sunspot observation taken after noon (note that it is the same day as in Table~\ref{tab:hv_rec_1}). Fig.~\ref{fig:fig2} illustrates the transits of the solar disk and the sunspot.

\begin{table}
\caption{Observation on 1796 July 07.}
\centering
\begin{tabular}{llll}
\hline
Object & Event & Recorded time \\
\hline
Sun &     h & 06:48:43 \\
Sun &     v & 06:49:48 \\
Spot &     h & 06:50:44 \\
Spot &     v & 06:51:48 \\
Sun &     h & 06:51:57 \\
Sun &     v & 06:53:06 \\
\hline
\end{tabular}
\label{tab:hv_rec_2}
\end{table}

For observations taken after noon, the inclination angle $i$ becomes positive, and 
in equation~\ref{eq:hv_xy}, the $\widetilde{y}$-coordinate becomes mirrored:
\begin{equation}
\begin{array}{l}
    \widetilde{y} = 2\delta_{\text{h}} / \Delta_{\text{h}} - 1 
\end{array}
\end{equation}

In particular, for the transits shown in Table~\ref{tab:hv_rec_2}, we obtain $\rho_\odot=69.28$~s, $i=45.58^{\circ}$,
$x=-0.33$,	and $y=0.02$, corresponding to a heliographic latitude of $b=5.2^{\circ}$ and a Carrington longitude of $L=285.9^{\circ}$. Comparing this result with the heliographic coordinates obtained from Table~\ref{tab:hv_rec_1}, we conclude that the two sets of measurements describe the same sunspot observed before and after noon on the same day. The result also indicates that we interpret the observational scheme correctly.

\begin{figure}
\includegraphics[width=\columnwidth]{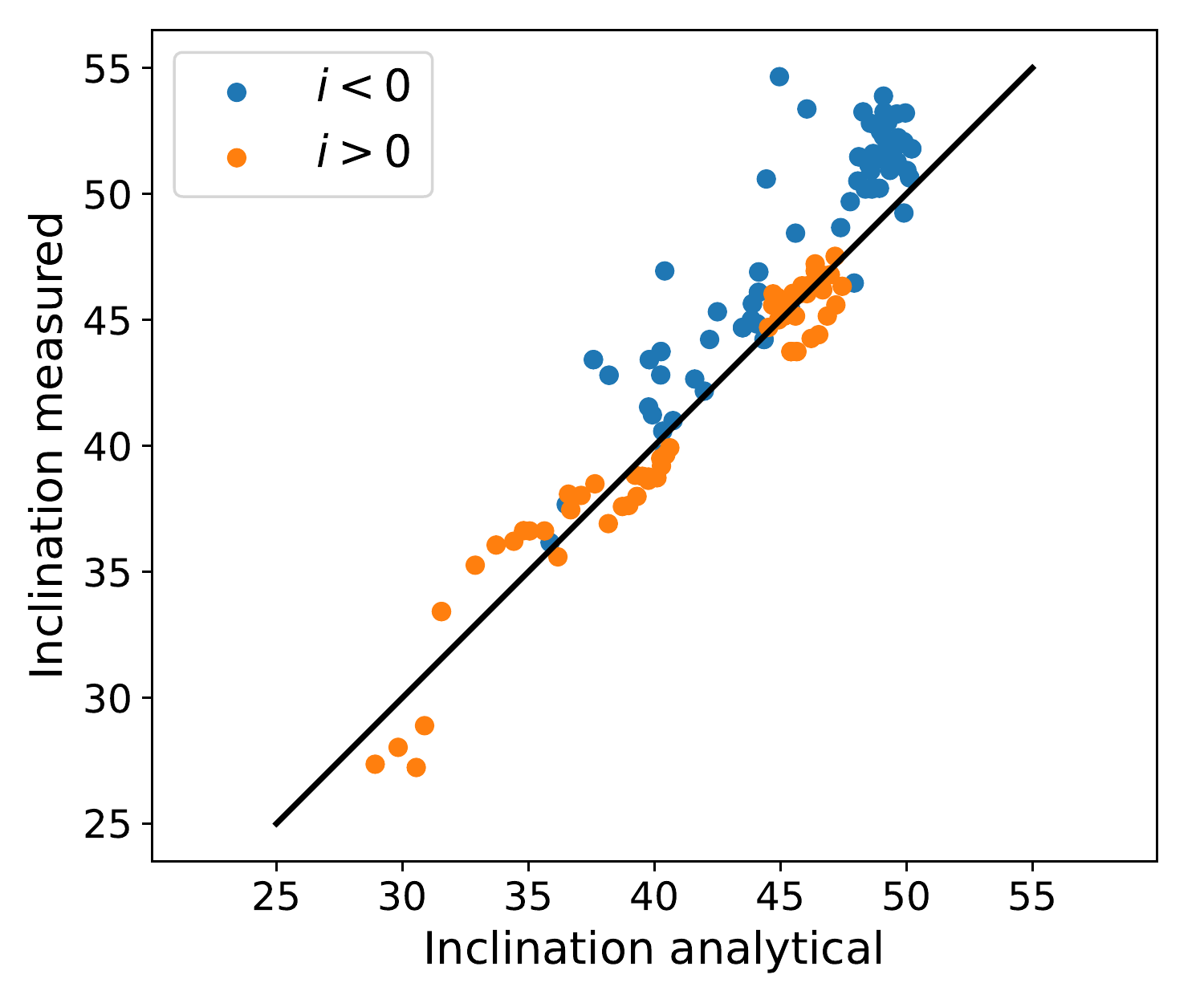}
    \caption{Comparison of the inclinations of the line of solar motion with the horizon, computed for the given time (analytical) and inferred from the observations (measured, $i$). The angles are absolute values, while blue and orange colors denote negative (before noon) and positive (after noon) signs, respectively.}
    \label{fig:incl}
\end{figure}

In contrast to Table~\ref{tab:hv_rec_1} and Table~\ref{tab:hv_rec_2} which allow the explicit reconstruction of sunspot coordinates, there are some incomplete observations where one or more transit events are missing (e.g. because of clouds). If $\Delta_{\text h}$ or $\Delta_{\text v}$ cannot be computed from the records, we compute the theoretical inclination angle $i$, based on the time and place of the observation, and use equation~\ref{eq:hv_alpha} to reconstruct the missing value. Fig.~\ref{fig:incl} shows that for complete observations there is a good correlation between the analytical inclination angles and angles derived from observations. 

\begin{figure}
\includegraphics[width=\columnwidth]{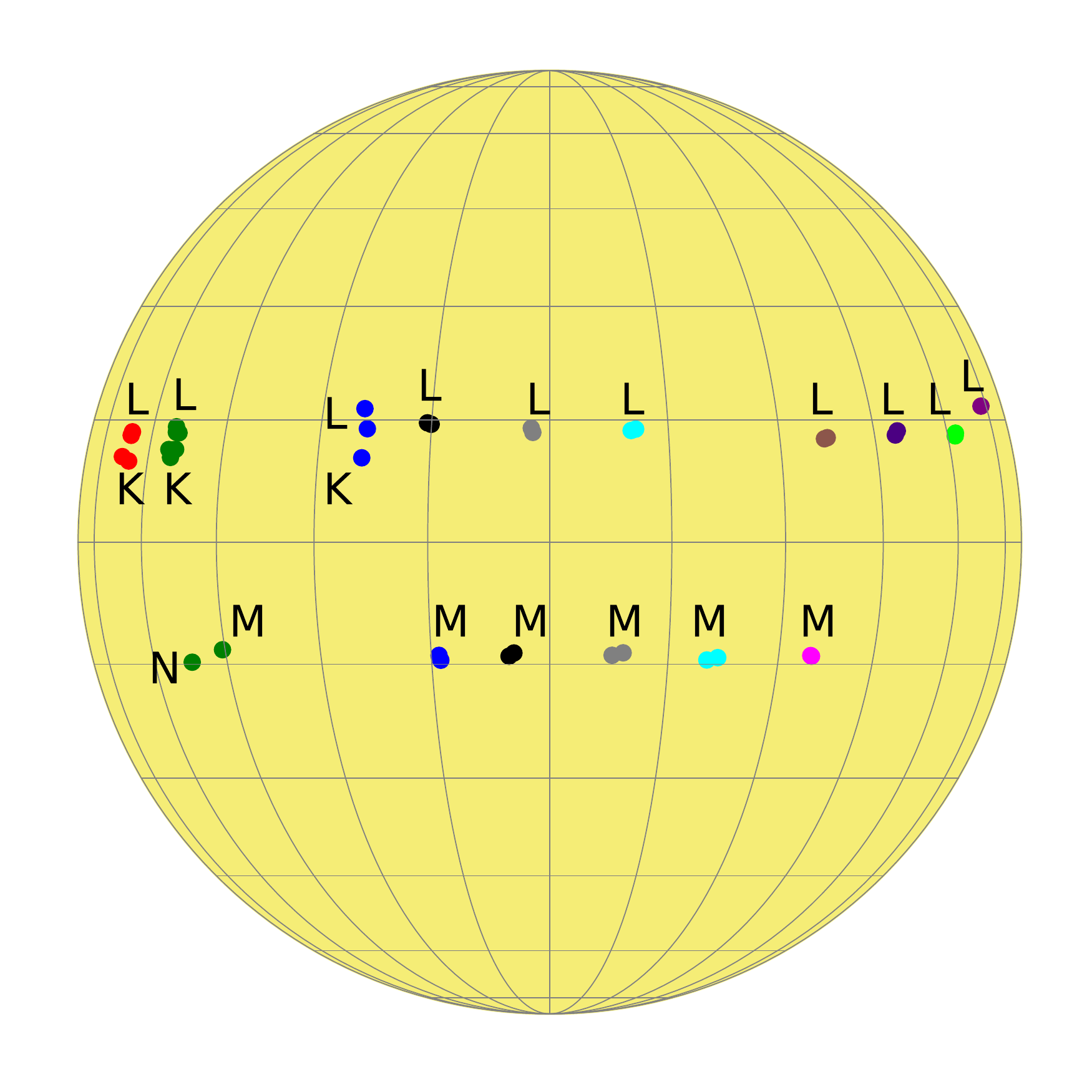}
    \caption{Sunspot observations during 1796 September~1--1796 September~12 in heliographic coordinates. Different colors correspond to different days. }
    \label{fig:hv_disk}
\end{figure}

Fig.~\ref{fig:hv_disk} shows the reconstructed heliographic coordinates for the period 1796 September~1 -- 1796 September~12.

Note that in the reconstruction procedure (\ref{eq:hv_deltas})--(\ref{eq:hv_xy}),
we approximated the solar motion linearly as a straight line. In order 
to estimate the possible error caused by this approximation, we 
calculated the absolute difference between the inclination angle $i$ at 
the time of first contact and at the time of last contact for each record.
We found that the average difference is about $0.5\degr$ and the maximum
value is $1\degr$. We assume that this effect does not dominate the other
inaccuracies in the measurements.

\subsection{Oblique wires}
In 1795--1800 and in 1806, Flaugergues also used a system with an `hour wire' which is a right ascension line, i.e.\ aligned with the equatorial coordinate system in contrast to the horizontal system of the previous Section. The hour wire is complemented by two mirror-symmetric oblique wires at which contact times are recorded as well. 
The symmetry follows from the fact that, for all records, 
the time interval between the contacts of the spot with the first 
oblique wire and the hour wire is approximately equal to the time 
interval between the contacts of the spot with the hour wire and 
the second oblique wire. The actual disagreement can be caused by 
telescope positioning errors, which we estimate below, and, less
importantly, by the curvature of the apparent motion of the Sun.

\begin{figure}
\includegraphics[width=\columnwidth]{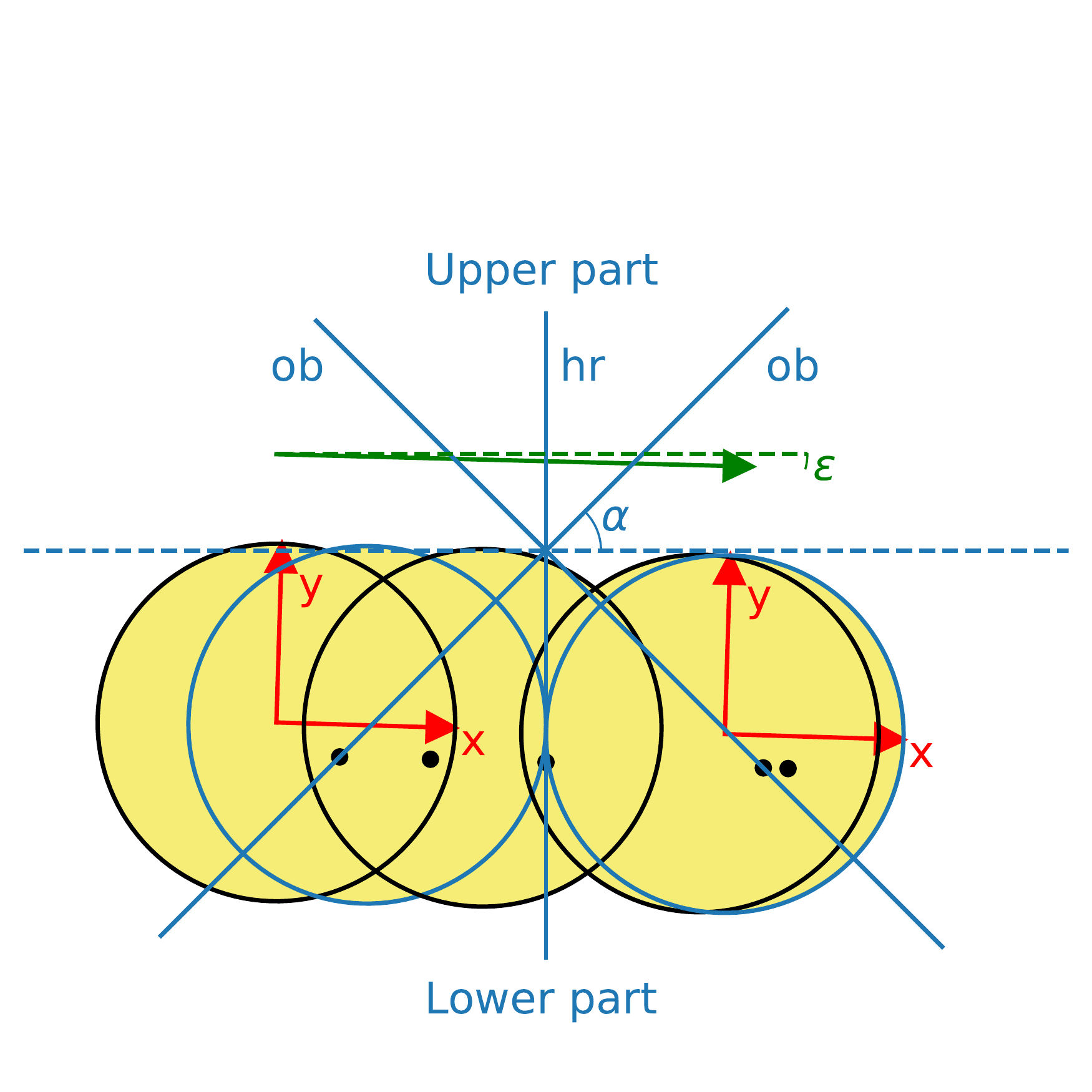}
    \caption{Sequence of transits of the solar disk and the sunspot through the set of oblique (ob) and hour (hr) wires corresponding to Table~\ref{tab:ob_rec_1}. The blue circles show contacts of the solar disk with the wires. The black circles show contacts of the sunspot with the wires. The dashed blue line is orthogonal to the hour line and represents the motion of the telescope. The green arrow is parallel to the line of solar motion and shows the direction of the motion. The angle between the motion of the telescope and the line of solar motion is denoted by $\varepsilon$ (note that $\varepsilon>0$ in this figure). The red arrows show the coordinate system aligned with the solar disk and the $x$-axis is parallel to the motion of the Sun.}
    \label{fig:fig3}
\end{figure}

Table~\ref{tab:ob_rec_1} shows an example of transit times obtained with this set of oblique and hour wires. Fig.~\ref{fig:fig3} illustrates this scheme. Note again that we plot the events assuming the direct motion of the Sun while Flaugergues could in fact have observed them upside down or mirrored. However, Flaugergues usually indicated which limb of the Sun (northern or southern) moves along the parallel line. This allows us to decide which part (upper or lower) should be considered. If no mention of the limb is given, we use the information from the previous and following days to obtain consistent coordinates.

The angle between the hour wire and the oblique wires is a free parameter. Below we will estimate it using an appropriate set of records and show that it is close to $45\degr$. At the moment we note that in \cite{fla}, pp. 321--322, we find a confirmation of this estimation (translated from French):
\begin{quote}`It does not appear that astronomers have, until now, used the rhomboid reticule to observe the spots of the Sun and the Moon. All the observations of this kind which are known to me, and for which reticules were used, were made by means of the passages of the spots and edges of the Sun and the Moon, or of the horns of that latter star, when it was in crescent, through the horizontal and vertical wires of the telescope of a quarter of a circle, or by the hour wire and the oblique ones of a reticule of $45^{\circ}$.'
\end{quote}

\begin{table}
\caption{Observation on 1795 March~6.}
\centering
\begin{tabular}{llll}
\hline
Object & Event & Recorded time \\
\hline
Spot &    ob & 12:49:39 \\
Sun &    hr & 12:50:12 \\
Spot &    hr & 12:50:54 \\
Spot &    ob & 12:52:13 \\
Sun &    hr & 12:52:22 \\
\hline
\end{tabular}
\label{tab:ob_rec_1}
\end{table}

As described above, if the time difference between the contacts of the sunspot with the first oblique wire and the hour line is not equal to the time difference between the contacts of the sunspot with the hour line and the second oblique wire, we need to estimate the positioning error of the telescope.

Let $\varepsilon$ be an inclination of the line of solar motion against the motion direction of the telescope. We assume that $\varepsilon$ is positive clockwise if the Sun moves in the lower part of oblique wires and positive counter-clockwise if the Sun is in the upper part.

Denote
\begin{equation}
\begin{array}{l}
    \Delta_{\text{1}} = t(\text{Sunspot at hr}) - t(\text{Sunspot at ob first time}) \\
    \Delta_{\text{2}} = t(\text{Sunspot at ob second time}) - t(\text{Sunspot at hr})
\end{array}
\label{eq:ob_Delte}
\end{equation}
Let $\alpha$ be the angle between either oblique wire and the motion line of the telescope (thus the angle between the hour wire and the oblique wire is $90\degr-\alpha$). It will turn out to be $\alpha=45\degr$ below.
It can be shown that
\begin{equation}
    \varepsilon = \arctan\left(\frac{\Delta_2-\Delta_1}{\Delta_1+\Delta_2}\tan\alpha\right) \, .
\label{eq:ob_err}
\end{equation}

If we denote
\begin{equation}
    \Delta = t(\odot\text{ at hr second time}) - t(\odot\text{ at hr first time}) \, ,
\label{eq:ob_delta_hr}
\end{equation}
then
\begin{equation}
    \rho_\odot = \frac{\Delta}{2}\cos\varepsilon \, .
\label{eq:ob_r}
\end{equation}

Then we compute a set of time differences for a sunspot:
\begin{equation}
\begin{array}{l}
    \delta_1 = t(\text{Sunspot at hr}) - t(\odot\text{ at hr first time}) \\
    \delta_2 = t(\text{Sunspot at ob first time}) - t(\odot\text{ at hr first time})
\end{array}
\label{eq:ob_spot}
\end{equation}
Note that $\delta_2 = \delta_1 - \Delta_1$ can be negative like in our example in Fig.~\ref{fig:fig3}.

\begin{figure}
\includegraphics[width=\columnwidth]{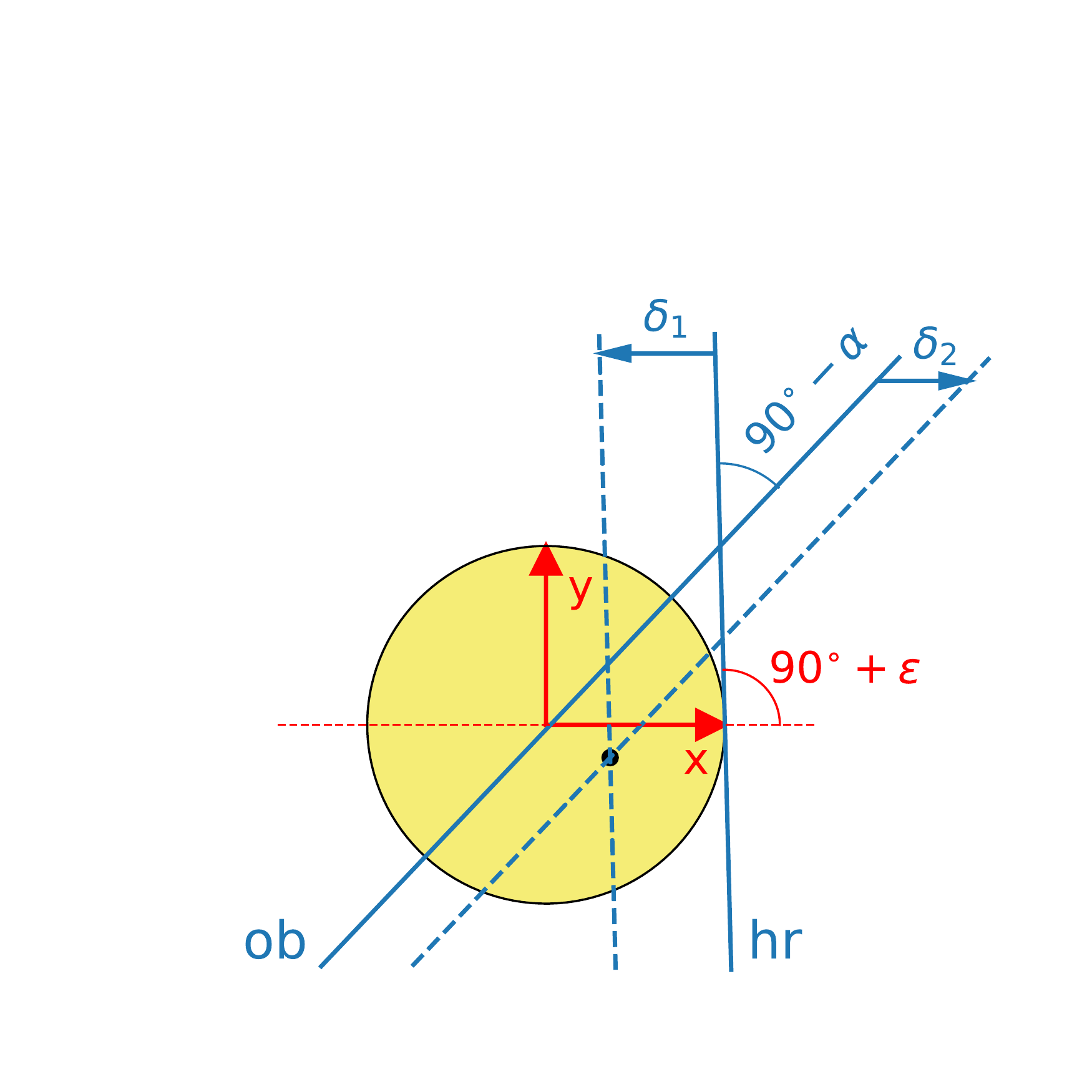}
    \caption{Towards the determination of a sunspot position in the lower part of oblique wires. The red arrows depict the coordinate system centred and normalised with the solar disk with the $x$-axis being parallel to the celestial equator. The solid blue lines show the positions of the hour wire and the first oblique wire at the time of first contact of the disk with the hour wire. The dashed blue lines show the positions of solid blue lines at two different times, respectively shifted by $\delta_1$ and $\delta_2$. The intersection of the dashed lines defines the position of the sunspot.}
    \label{fig:ob_coord}
\end{figure}

Now it is convenient to sit in a coordinate system moving with the solar disk. In this coordinate system, we observe wires passing through the disk. The position of the sunspot is defined by the intersection of the wires that moved $\delta_1$ and $\delta_2$ seconds from the first contact of the solar disk with the hour line (see Fig.~\ref{fig:ob_coord} for a pictorial explanation).

Taking into account that $-1\le x\le1$ and $-1\le y\le1$, we obtain a system of linear equations which defines the position of the sunspot:
\begin{equation}
\begin{array}{l}
    (y + c)\rho_\odot = c\tan(90^{\circ}+\varepsilon)\left(x\rho_\odot - \frac{\rho_\odot}{\tan((90^{\circ}+\varepsilon)/2)} + \delta_1 \right)  \\
    (y + c)\rho_\odot = -c\tan(\alpha+\varepsilon)\left(x\rho_\odot - \frac{\rho_\odot}{\tan((90^{\circ}+\varepsilon)/2)} + \delta_2 \right)  \, ,
\end{array}
\label{eq:ob_system}
\end{equation}
where $c=1$ if the Sun is in the upper part of the oblique-wire system and $-1$ otherwise.

The solution is
\begin{equation}
    x = \frac{1}{\tan((90^{\circ}+\varepsilon)/2)} - 
    \frac{1}{\rho_\odot}\frac{\delta_1\tan(90^{\circ}+\varepsilon)+
    \delta_2\tan(\alpha+\varepsilon)}{\tan(90^{\circ}+\varepsilon) + \tan(\alpha+\varepsilon)}  \, ,
\label{eq:ob_x}
\end{equation}
and $y$ can be derived from the first or second equation in \ref{eq:ob_system}.

Note that for $\varepsilon=0$ the solution simplifies to
\begin{equation}
\begin{array}{l}
    x = 1 - \frac{\delta_1}{\rho_\odot}  \\
    y = -c + c\frac{\delta_1 - \delta_2}{\rho_\odot}\tan\alpha  \, ,
\end{array}
\label{eq:ob_xy}
\end{equation}

Finally, using the solar parameters $B_0$, $L_0$, and $P$ for the time of observation, we map $x$ and $y$ to heliographic coordinates.  

For the transits shown in Table~\ref{tab:ob_rec_1} in particular, we obtain $\varepsilon=1.49$,
$\rho_\odot=64.98~s$, 
$x=0.36$, and $y=-0.18$. They correspond to a heliographic latitude of $b=-8.1^{\circ}$ and a Carrington longitude of $L=295.4^{\circ}$.

It is possible that either the first or the second transit of a sunspot at the oblique wires is missing. In this case we assume that $\varepsilon=0$ and $\Delta_1=\Delta_2$. We obtain $\delta_2$ from $\delta_2=\delta_1-\Delta_1$ or $\delta_2=\delta_1-\Delta_2$.

Given the same spot was recorded by sets of transits in both the upper and lower parts of the wire system, we can estimate the actual value of the angle $\alpha$. We consider two observations of the same spot made on 1795 March~4 in different parts of the field of view (see Table~\ref{tab:ob_rec_alpha}). For simplicity, we neglect the positioning error (i.e.\ assume $\varepsilon=0$) and use the second equation in~(\ref{eq:ob_xy}) to estimate~$\alpha$.

\begin{table}
\caption{Two observations of the same sunspot on 1795 March~4. The first set of timings is made in the lower part of the oblique wires, the second in the upper part. The sunspot is labeled as \textit{grande} (large) in the original records.}
\centering
\begin{tabular}{lllll}
\hline
Object & Event & Transit times & Transit times\\
       &       & in lower part & in upper part\\
\hline
Sun &    hr & 12:55:34 & 01:01:58.5 \\
Spot &    ob & 12:56:00 & 01:02:07 \\
Spot &    hr & 12:56:57 & 01:03:21 \\
Sun &    hr & 12:57:45 & 01:04:09 \\
Spot &    ob & 12:57:55 & 01:04:30 \\
\hline
\end{tabular}
\label{tab:ob_rec_alpha}
\end{table}

Indeed, for the first record in Table~\ref{tab:ob_rec_alpha} we obtain $\delta_1=83$~s, $\delta_2=26$~s, and $\rho_\odot=65.5$~s, while for the second record $\delta_1=82$~s, $\delta_2=8$~s, and $\rho_\odot=65$~s. Substituting these values in equation~\ref{eq:ob_xy} and taking into account that the $y$-coordinate is equal in both cases, we obtain $\tan\alpha=1.004$ and conclude that $\alpha=45\degr$.

The positioning error was estimated using equation~(\ref{eq:ob_err}) and was found to be usually $|\varepsilon|<3\degr$.

\begin{figure}
\includegraphics[width=\columnwidth]{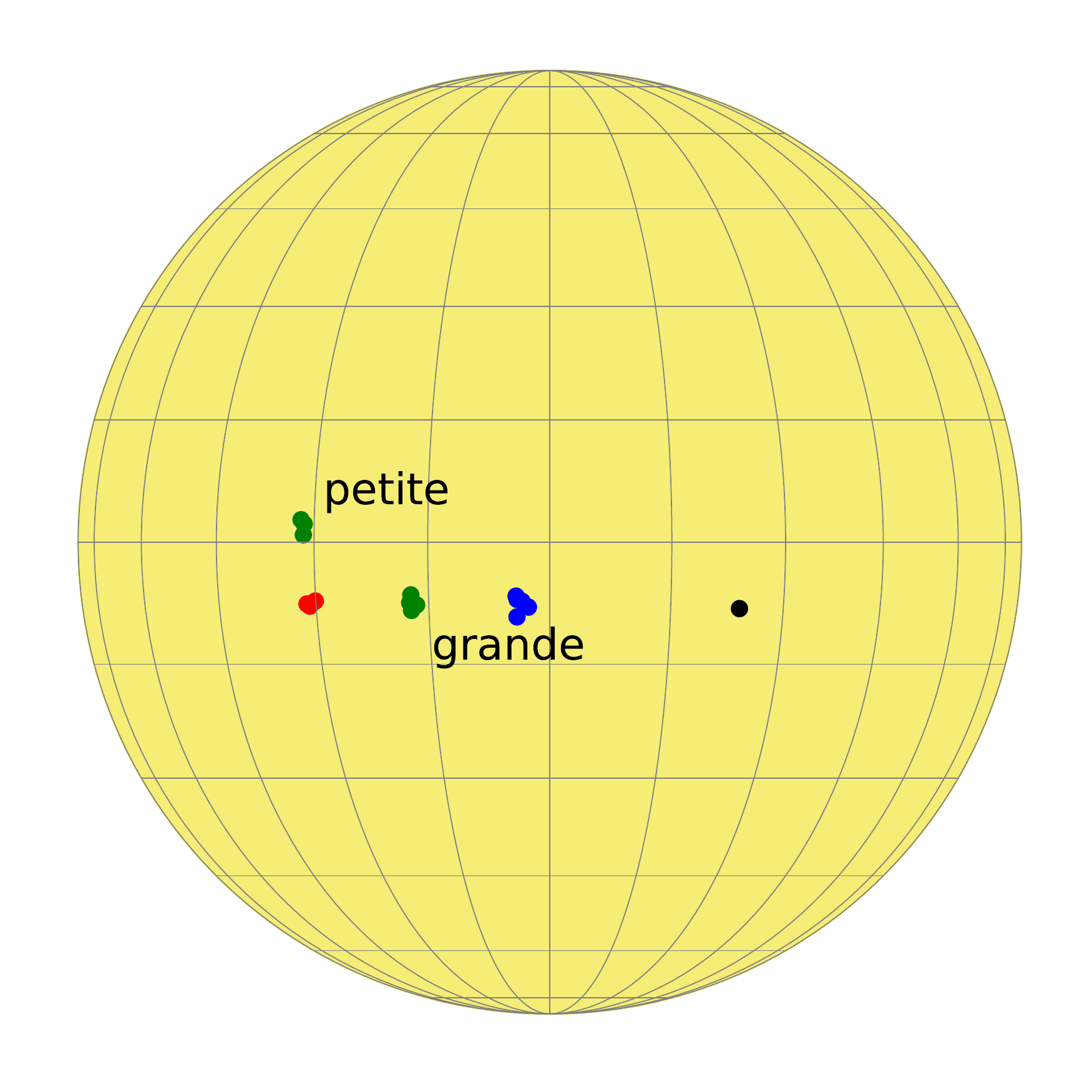}
    \caption{Sunspot locations during 1795 March~3--1795 March~6 (actually most likely March~7) in heliographic coordinates. Different colors correspond to different days in the order red, green, blue, and black. The two sunspots observed on 1795 March~4 are labeled as \textit{grande} (large) and \textit{petite} (small). }
    \label{fig:ob_disk}
\end{figure}

In Fig.~\ref{fig:ob_disk} we show the reconstructed sunspot positions for 
the period 1795 March~3--1795 March~6. Since there is a change in the central
meridian distance of almost $30\degr$ between the last two days, it is
very likely that the last date was actually 1795 March~7. In fact,
Flaugergues corrected his notes from March~7 to March~6 and from 
March~8 to March~7, and continues
with March~9, while March~8 is missing. His correction may have been
wrong, or made according to the astronomical date which starts at noon
of the civil date of the same numerical value (the astronomical
calendar is therefore `late' by 12~hours). So the astronomical March~7
may not have had started, and Flaugergues changed to March~6.

We find one day (1796 October~14) when Flaugergues made observations both with oblique wires and cross-hair (horizontal and vertical) wires, apparently for the same sunspot. 
This allows us to verify the coordinate reconstruction methods.
The two measurements with orthogonal wires yield $b=3.6\degr$, $L=23.5\degr$ and $b=3.3\degr$, $L=23.8\degr$. The two measurements with oblique wires yield $b=4.1\degr$, $L=25.4\degr$ and $b=3.3\degr$, $L=24.8\degr$. Since both methods show similar coordinates, we conclude that the same sunspot was observed and we correctly reconstructed the coordinates.

\subsection{Rhombus}
\label{sec:rhomb}

\begin{figure}
\includegraphics[width=\columnwidth]{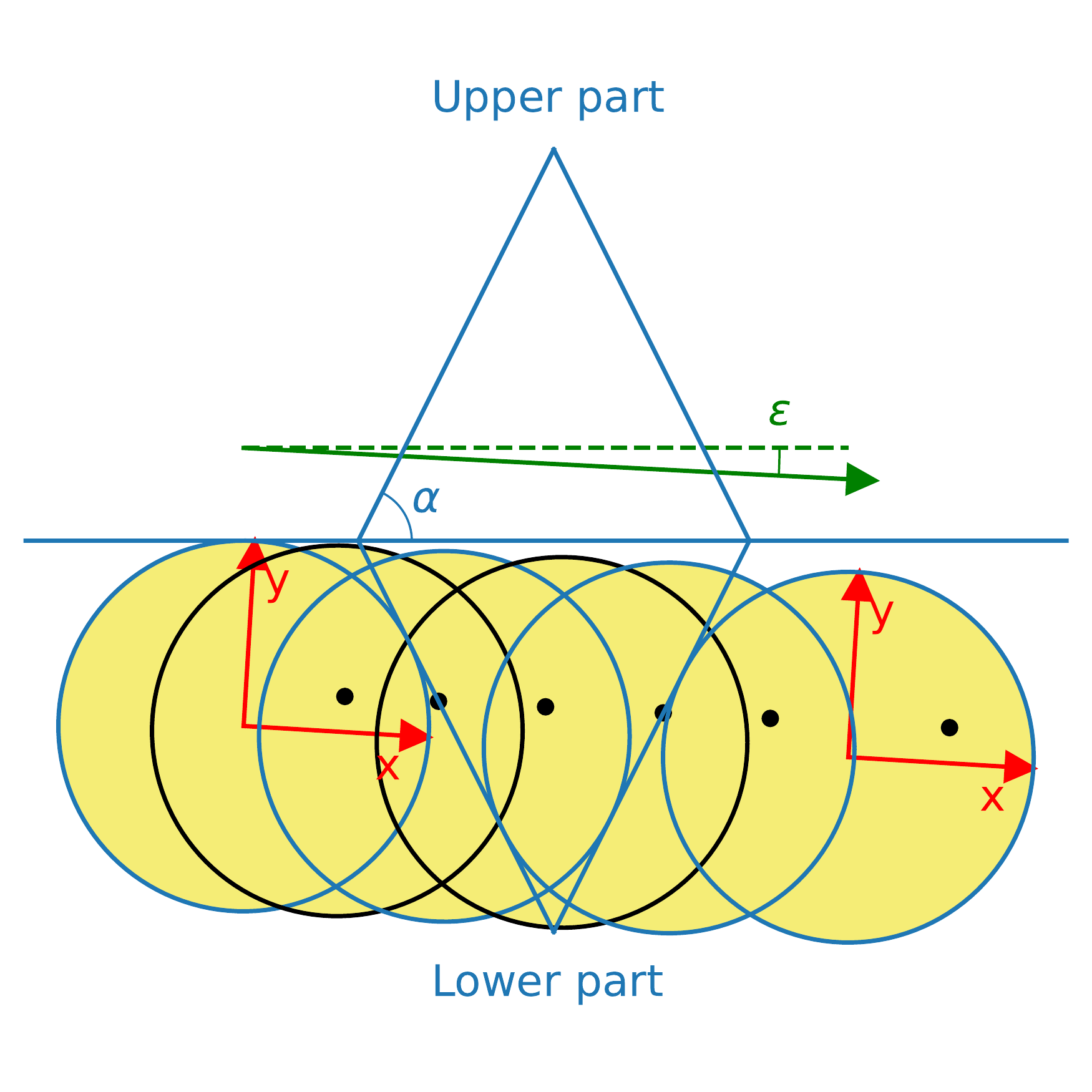}
    \caption{Sequence of transits of the solar disk and the sunspot through the rhombus corresponding to the Table~\ref{tab:rh_rec_1}. The blue circles show contacts of the solar disk with the wires. The black circles show contacts of the sunspot with the wires.
    The green arrow again shows the direction of the solar motion. The angle between the short diagonal of the rhombus and the line of solar motion is denoted by $\varepsilon$ (note that $\varepsilon>0$ in this Figure). The red arrows show the coordinate system aligned
    with the solar disk with the $x$-axis being parallel to the apparent solar motion.}
    \label{fig:fig4}
\end{figure}

In this observing scheme used in 1813--1830, a set of wires form a rhombus and the Sun moves through the upper or lower part of the rhombus (see Fig.~\ref{fig:fig4}). During the rotation of the Earth, the leading part of the solar disk touches the eastern side of the rhombus on the outside (contact 1), then the leading part of the solar disk touches the western side of the rhombus on the inside  (contact 2), then the trailing part of the solar disk touches the eastern side of the rhombus on the inside (contact 3), and finally the trailing part of the solar disk touches the western side of the rhombus on the outside (contact 4). Between the first and the last contacts, the sunspot touches the eastern and western sides of the rhombus. Table~\ref{tab:rh_rec_1} shows an example of transit times which correspond to Fig.~\ref{fig:fig4}.

Note that Fig.~\ref{fig:fig4} shows the direct motion of the Sun, while Flaugergues observed the upside-down image in fact (we find a remark on 1816 October~5 that supports this proposition). This should be taken into account when interpreting the sketches in his manuscript, which indicate in which part of the rhombus the Sun was observed.

\begin{table}
\caption{Observation on 1830 July~2.}
\centering
\begin{tabular}{llll}
\hline
Object & Event & Recorded time \\
\hline
Sun &    contact 1 & 02:41:24 \\
Spot &    in & 02:41:59.5 \\
Sun &    contact 2 & 02:42:39 \\
Spot &   out & 02:43:23 \\
Sun &    contact 3 & 02:44:03 \\
Sun &   contact 4 & 02:45:10 \\
\hline
\end{tabular}
\label{tab:rh_rec_1}
\end{table}

The angle $\alpha$ in the basement of the rhombus is a free parameter. \cite{fla} compares different types of rhombuses (reticules) regarding their accuracy as well as their practicability. While he noticed that most solar and lunar observations had been made with the Lalande\footnote{J\'er\^ome Lalande (1732--1807), a French astronomer and director at Paris Observatory in 1795--1807 \citep{Lalande}.} type of reticules of $\alpha=45\degr$ for the oblique wires, he actually compares the rhombus by Bradley\footnote{James Bradley (1692--1762), an English astronomer and third Astronomer Royal at Greenwich \citep{Bradley}.}  with $\alpha=\arctan 2 \approx 63.4\degr$, and one made of equilateral triangles with $\alpha=60\degr$. Bradley's reticule results from a rhombus whose long diagonal is twice as long as the short diagonal. Flaugergues does not say which one he actually used, but he wraps up his considerations with 

\begin{quote}
`The calculation of the difference of declinations is a little simpler, using the Bradley reticule; but this small advantage does not compensate for the difficulty of constructing exactly this reticule. I propose to astronomers to substitute the rhomboid chosen by Bradley with a rhomboid composed of two opposite equilateral triangles described on the same line serving as a base, which becomes the small diagonal of this rhomboid represented in the figure. The description of the equilateral triangle, which is the subject of the first proposition of Euclid's elements, is the simplest and easiest of all geometrical operations. One will thus be able to construct, with the greatest exactitude, the reticule that I propose; ...'
\end{quote}

Although $\arctan 2 \approx 63.4\degr$ is close to $60\degr$, we will try to estimate $\alpha$ from the records by minimizing the latitudinal drift of the sunspots between neighbouring days. 

Normally, Flaugergues tried to align the rhombus so that the Sun moves in parallel to the base of the lower or the upper part of the rhombus. In this ideal case, the time difference between contact~1 and contact~3  equals the time difference between contact~2 and contact~4. Due to errors in the instrument alignment, however, the two time differences become unequal (we assume that the line of solar motion is a straight line). For example, in Table~\ref{tab:rh_rec_1} they are 159 and 151~seconds respectively. Our first goal is to determine the inclination of the line of the solar motion against the base of the lower or the upper part of the rhombus. 

Let us assume that the Sun crosses the lower part of the rhombus as it is shown in Fig.~\ref{fig:fig4} and let $\varepsilon$ be the angle between the base of the lower part of the rhombus and the line of the solar motion. We assume that the angle $\varepsilon$ is positive clockwise in the lower part and positive counter-clockwise in the upper part. Let
\begin{equation}
\begin{array}{l}
    \Delta_1 = t(\odot\text{ contact 3}) - t(\odot\text{ contact 1}) \\
    \Delta_2 = t(\odot\text{ contact 4}) - t(\odot\text{ contact 2}) \, .
\end{array}
\label{eq:rh_cont1}
\end{equation}
Then for the lower and upper parts of the rhombus
\begin{equation}
\begin{array}{l}
    2\rho_\odot = \Delta_1\sin(\alpha - \varepsilon) \\
    2\rho_\odot = \Delta_2\sin(\alpha + \varepsilon) \, ,
\end{array}
\label{eq:rh_rad}
\end{equation}
where $\alpha$ is the angle of the inclined triangle legs with the base of the lower and upper part of the rhombus.

From equation~\ref{eq:rh_rad} we obtain 
\begin{equation}
    \varepsilon = \arctan\left(\frac{\Delta_1 - \Delta_2}{\Delta_1 + \Delta_2}\tan\alpha\right)\, .
\label{eq:rh_err}
\end{equation}

Let us denote
\begin{equation}
\begin{array}{l}
    \delta_1 = t(\text{sunspot in}) - t(\odot\text{ contact 1}) \\
    \delta_2 = t(\odot\text{ contact 4}) - t(\text{sunspot out}) \, .
\end{array}
\label{eq:rh_cont2}
\end{equation}

\begin{figure}
\includegraphics[width=\columnwidth]{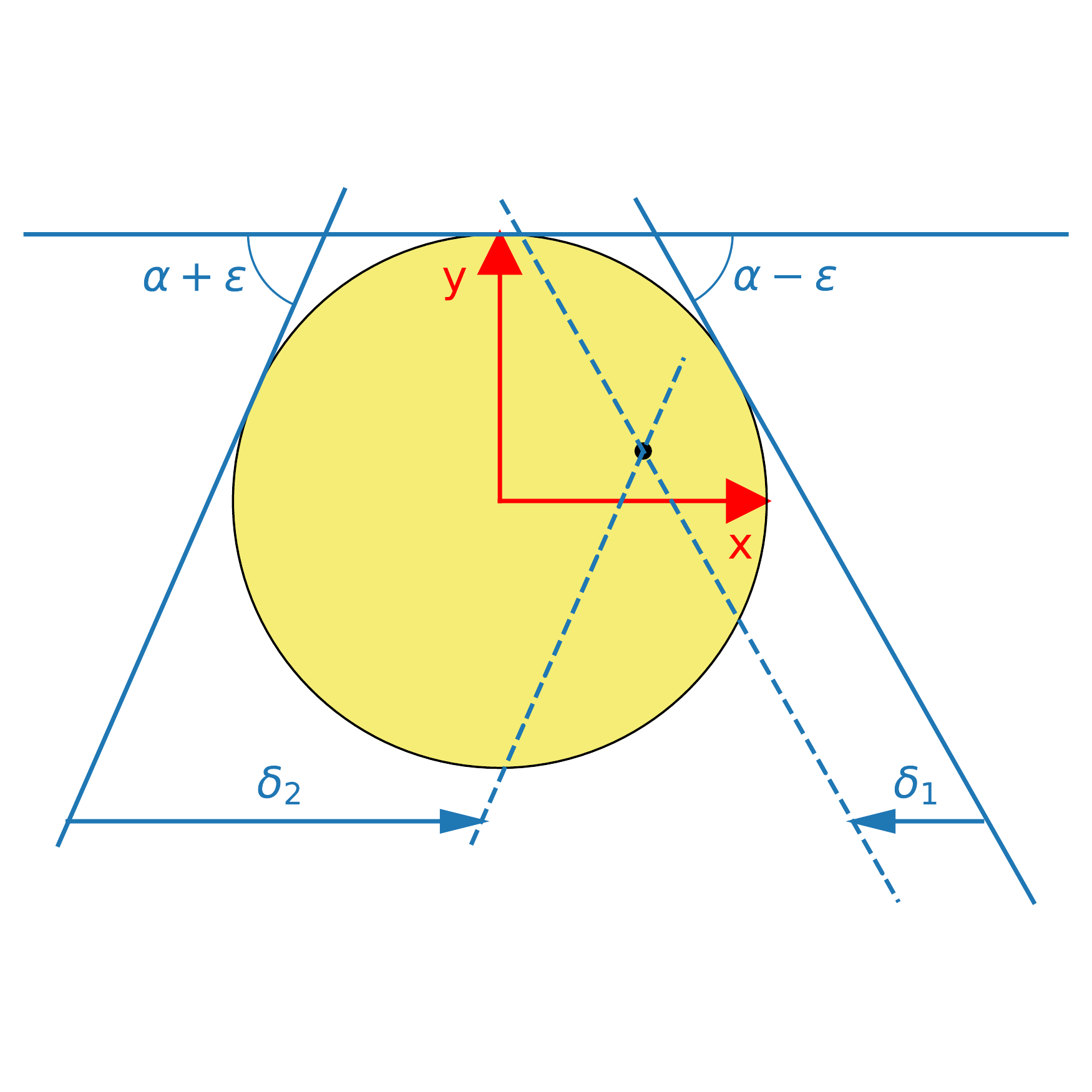}
    \caption{Towards the determination of sunspot coordinates in the lower part of the rhombus. The red arrows show the coordinate system aligned with the solar disk with the $x$-axis being parallel to the line of the apparent solar motion. The inclined blue lines show the two sides of the rhombus at the first and last external contacts with the solar disk. The dashed lines show the two sides of the rhombus shifted by $\delta_1$ and $\delta_2$ in time. The intersection of the dashed lines defines the position of the sunspot.}
    \label{fig:rh_coord}
\end{figure}

\begin{figure}
\includegraphics[width=\columnwidth]{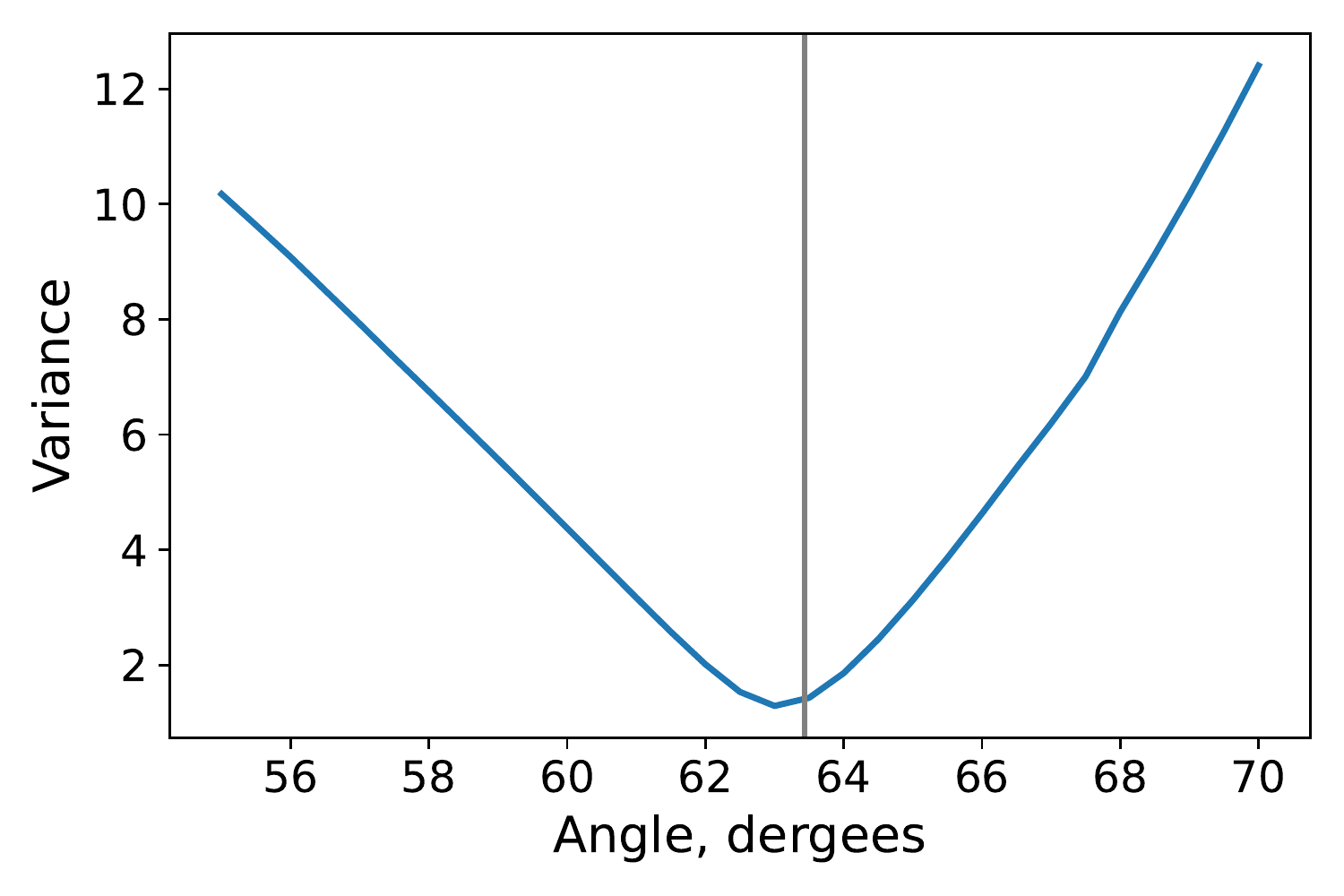}
    \caption{Sum of the latitudinal variances of several long sunspot series against the angle $\alpha$. The vertical gray line corresponds to $\arctan 2 $. }
    \label{fig:alpha}
\end{figure}

Now we sit in a coordinate system moving with the solar disk and observe wires passing across the disk (see Fig.~\ref{fig:rh_coord}). Taking into account that $-1\le x\le1$ and $-1\le y\le1$, we obtain a system of linear equations which defines the position of the sunspot as the intersection of the two lines:
\begin{eqnarray}
    (y+c)\rho_\odot &=& \phantom{-}c\tan(\alpha - 
                        \varepsilon)\left(x\rho_\odot -
                        \rho_\odot\tan\frac{\alpha-\varepsilon}{2} + \delta_1\right) \nonumber\\
    (y+c)\rho_\odot &=& -c\tan(\alpha + \varepsilon)\left(x\rho_\odot + \rho_\odot\tan\frac{\alpha+\varepsilon}{2} - \delta_2\right),\nonumber\\
                    & & 
\label{eq:rh_system}
\end{eqnarray}
where $c=1$ in the upper part and $c=-1$ in the lower part of the rhombus.

The solution is
\begin{eqnarray}
    x &=& \frac{1}{\tan(\alpha - \varepsilon)+\tan(\alpha + \varepsilon)}\times \notag\\
      & & \Biggl[ \tan(\alpha - \varepsilon)\left(\tan\frac{\alpha-\varepsilon}{2} - \frac{\delta_1}{\rho_\odot}\right) - \notag\\
      & & \tan(\alpha + \varepsilon)\left(\tan\frac{\alpha+\varepsilon}{2} - \frac{\delta_2}{\rho_\odot}\right)\Biggr]
\label{eq:rh_x}
\end{eqnarray}
and the $y$-coordinate can be easily derived from equation~(\ref{eq:rh_system}).

In some observations, Flaugergues skipped the internal contacts~2 and~3 and noted only the external contacts~1 and~4. In such cases, we cannot estimate $\varepsilon$ and assume $\varepsilon=0$. Without internal contacts we cannot obtain $\rho_{\odot}$ either and have to assume a constant value of $\rho_{\odot}=65$~s. Note that for $\varepsilon=0$ the solution simplifies to
\begin{eqnarray}
    x &=& \frac{\delta_2-\delta_1}{2\rho_\odot} \nonumber\\
    y &=& -c + c\tan\alpha\left(\frac{\delta_1+\delta_2}{2\rho_\odot} - \tan\frac{\alpha}{2}\right) \, .
\label{eq:rh_sol_simplt}
\end{eqnarray}

\begin{figure}
\includegraphics[width=\columnwidth]{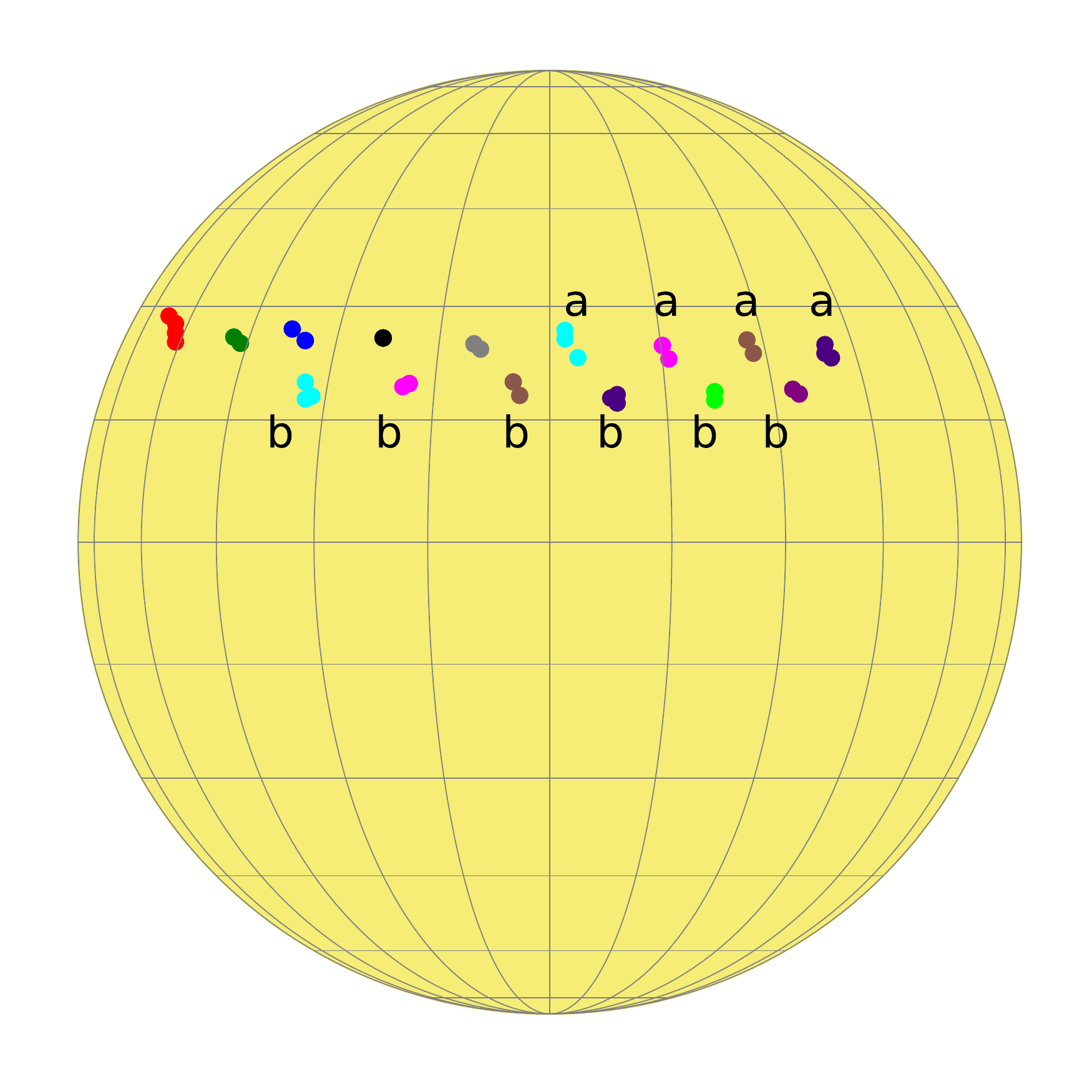}
    \caption{Sunspot observations during 1813 April~4--1813 April~14 in heliographic coordinates. Different colors correspond to different days.}
    \label{fig:rh_disk}
\end{figure}

In order to estimate $\alpha$, we select a number of series of sunspot observations of five days and more. Then we vary $\alpha$ from $55^{\circ}$ to  $70^{\circ}$ and compute the sum of the latitudinal variances of each spot. Fig.~\ref{fig:alpha} shows that the minimum variance is near $\arctan 2$ and is much less compatible with $60\degr$, corresponding to a rhombus of equilateral triangles. Hence, we use $\alpha = \arctan 2$ for the reconstruction of the coordinates. Fig.~\ref{fig:rh_disk} shows the reconstructed heliographic coordinates for the period 1813 April~4--1813 April~14. This is one of the longest series of consecutive observations of the same group of spots.

An estimation of the positioning errors using equation~\ref{eq:rh_err} gave in most cases $|\varepsilon|<3^{\circ}$.

\subsection{Single transit}
\label{sec:st}

Near noon, Flaugergues regularly measured transits of the solar disk and sunspots through a single wire (see Table~\ref{tab:m_rec} for example). Assuming that the line of the apparent solar motion is orthogonal to this wire at the moment of observation, one can reconstruct the $x$-coordinate of the sunspot with respect to the line of solar motion.

\begin{table}
\caption{Observation on 1830 March~22.}
\centering
\begin{tabular}{ll}
\hline
Object &  Recorded time \\
\hline
Sun &     0:12:20 \\
Spot &    0:13:41 \\
Sun &     0:14:30 \\
\hline
\end{tabular}
\label{tab:m_rec}
\end{table}

\begin{figure}
\centering
\includegraphics[width=0.7\columnwidth]{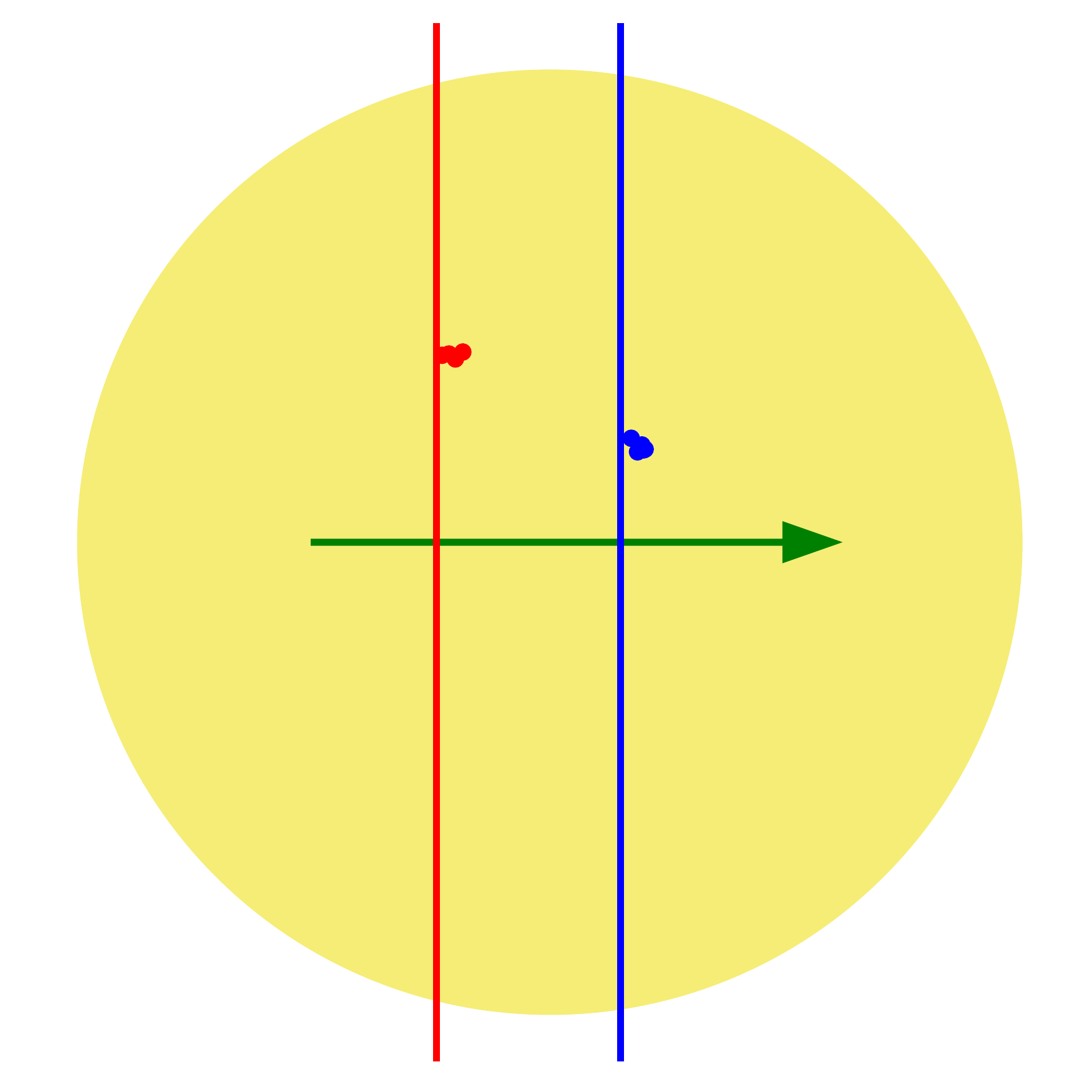}
    \caption{Sunspot positions reconstructed from measurements made with vertical and horizontal wires (dots) and from single transits (lines). Red colour corresponds to 1830 March~22 and blue colour to 1830 March~24. Note that here we use a coordinate system where the $x$-axis represents the line of solar motion (green arrow).}
    \label{fig:m_disk}
\end{figure}

We find several days when simple transits were accompanied by additional measurements allowing the reconstruction of both coordinates. In particular, on 1830 March~22 and on 1830 March~24 there were measurements made with horizontal and vertical wires (see Section~\ref{sec:hv}) and single-transit measurements. The dots in Fig.~\ref{fig:m_disk} show sunspot positions reconstructed from measurements made with horizontal and vertical wires while vertical lines show possible sunspot positions that correspond to single transits. Since we find that the sunspots are close to the vertical lines, our interpretation of both measurement systems is confirmed.

\begin{figure}
\centering
\includegraphics[width=0.7\columnwidth]{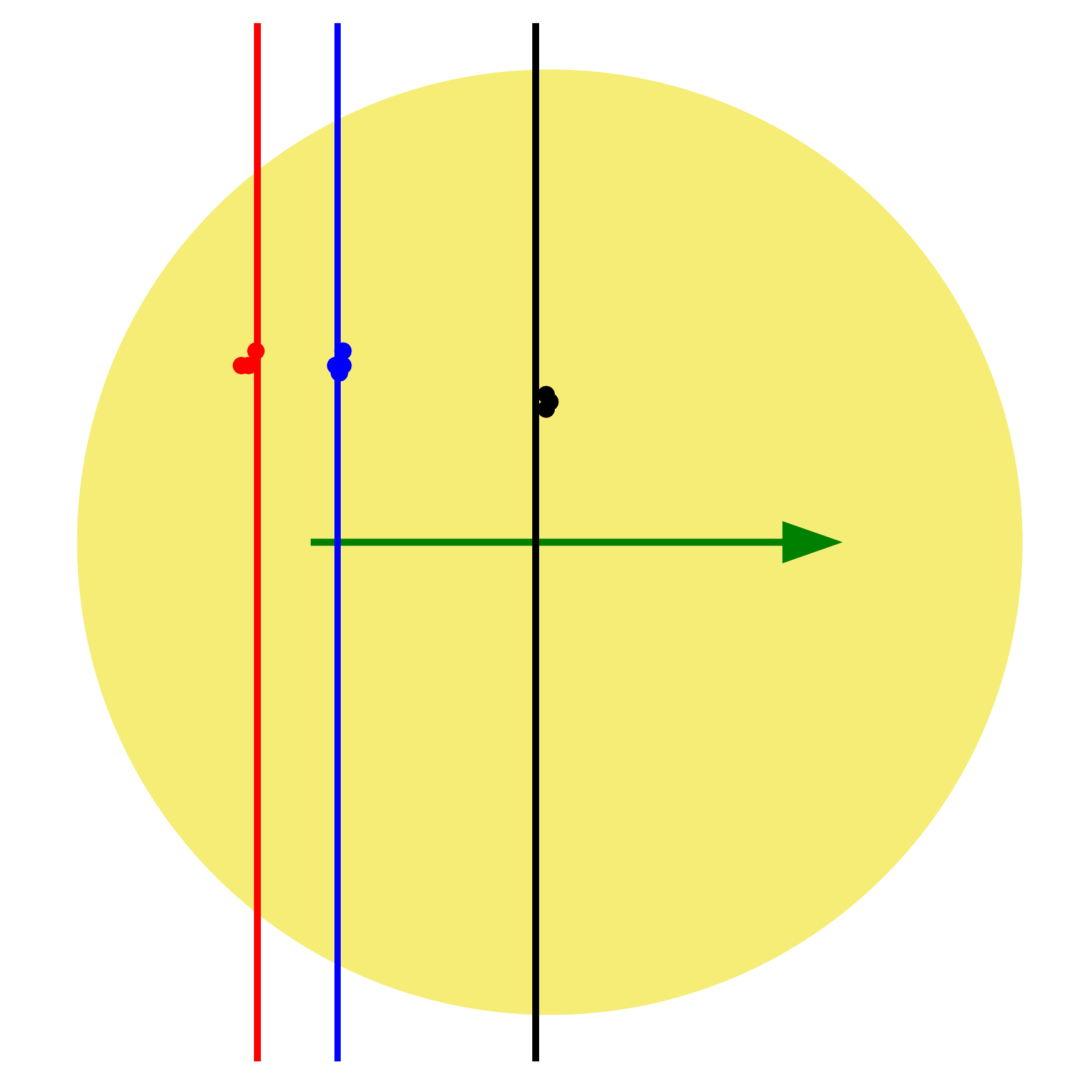}
    \caption{Sunspot positions reconstructed from measurements with the rhombus (dots) and from single transits (lines). Different colours correspond to different days (red for 1813 February~3, blue for 1813 February~4, and black for 1813 February~6). Again, the $x$-axis represents the line of solar motion (green arrow).}
    \label{fig:m2_disk}
\end{figure}

Similarly, Fig.~\ref{fig:m2_disk} shows sunspot positions reconstructed from measurements made with the rhombus (see sec.~\ref{sec:rhomb}) while vertical lines represent the possible sunspot positions corresponding to single transits. Different colours correspond to different days in the period 1813 February~3--1813 February~6.

We do not find any days for which both single transits and measurements with oblique wires were recorded.

We use single transit measurements for the additional verification of reconstructed coordinates or as an indicator for the presence of sunspots on specific days.

\subsection{Three-wire scheme}

There is one more observational setup for which, however, we do not have a clear interpretation. It is first mentioned on 1817 February~19, and contains solar-disk and sunspot transits through a set of three wires as shown in Table~\ref{tab:3w_rec}. Usually, we find these measurements near noon. Given the time differences between the transits, it appears that all wires are
parallel to the middle one and are placed at equal distances. We suppose that the small remaining time difference between transits from the first to the second wire and from the second to the third wire are probably due to the fuzziness of the solar limb and the subjectivity in fixing the actual transit time.

As a result, only the coordinate along the solar motion can be estimated. Like in the case of single transits (sec.~\ref{sec:st}), we use these measurements for additional verification of coordinates, reconstructed from other measurements, or for detecting the presence of sunspots on specific days. 

\begin{table}
\caption{Observation on 1824 September~27. Notation of the events: f1 -- primum filum, fm -- filum meridianum, f3 -- tertium filum.}
\centering
\begin{tabular}{lll}
\hline
Object &  Event & Recorded time \\
\hline
     Sun &    f1 &  23:48:45 \\
     Sun &    fm &  23:49:12.75 \\
     Sun &    f3 &  23:49:44 \\
     Spot &    f1 &  23:49:35 \\
     Spot &    fm &  23:50:03 \\
     Spot &    f3 &  23:50:33 \\
     Sun &    f1 &  23:50:54.5 \\
     Sun &    fm &  23:51:21.75 \\
     Sun &    f3 &  23:51:52 \\
\hline
\end{tabular}
\label{tab:3w_rec}
\end{table}

\subsection{Solar eclipses}

During the two solar eclipses, on 1788 June~4 and 1816 Nov~19, Flaugergues recorded times when the disk of the Moon contacts the sunspots. This allows the reconstruction of sunspot coordinates. We start with the latter eclipse, since Flaugergues also measured the sunspot position with the rhomboid scheme (Sec.~\ref{sec:rhomb}) on that day, so we can compare the results of both methods. We use the {\sc astropy} \citep{astropy:2013, astropy:2018, astropy:2022} package to reconstruct the solar eclipse.

Table~\ref{tab:eclipse1816} contains the solar eclipse contact times recorded on 1816 November~19. 
Using {\sc astropy}, we estimated the start time of the solar eclipse as $8^{\rm h}04^{\rm m}15^{\rm s}$ and the end time as $10^{\rm h}23^{\rm m}39^{\rm s}$ (both times are UTC and are valid for the observing site).
Note these times are just estimates made with {\sc astropy}, but the exact determination of an eclipse time
is a separate problem including many specific details \citep[see, e.g., ][]{Stephenson}. Since our coordinate derivation is relative to the observed contacts, the absolute timing of the eclipse is of lesser importance here.

\begin{figure}
\centering
\includegraphics[width=0.9\columnwidth]{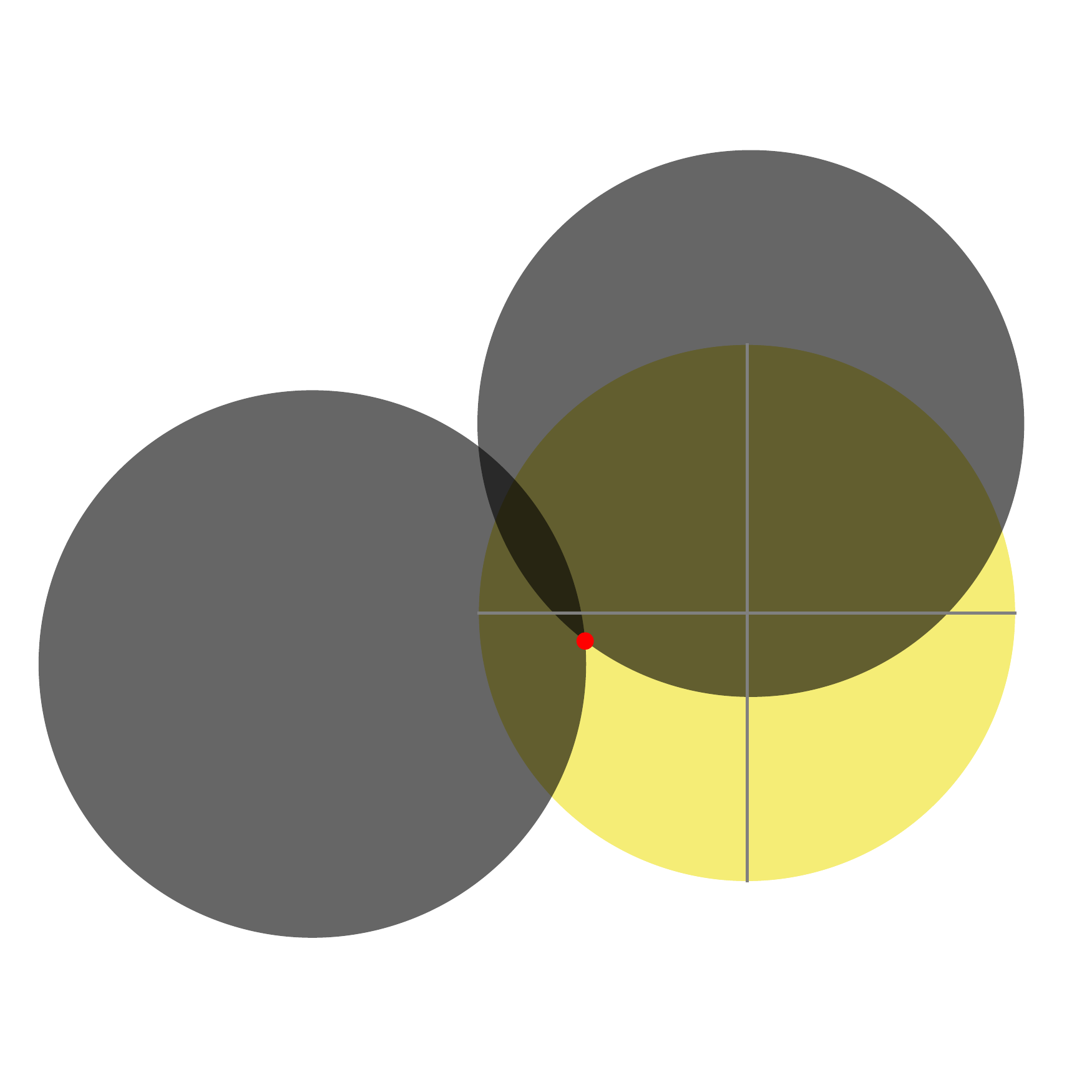}
    \caption{Apparent positions of the Sun and the Moon at the times when the Moon contacts the sunspot during the solar eclipse of 1816 November~19. The red point at the intersection of the lunar disks gives the position of the sunspot. The vertical direction in the plot points to the celestial north. The Moon moves from right to left with time.}
    \label{fig:2moons}
\end{figure}

Remarkably, the estimated duration of the eclipse is only two seconds longer than the historical records indicate. We therefore used the estimated start time of the eclipse to obtain UTC times of the recorded events (see Table~\ref{tab:eclipse1816}). Fig.~\ref{fig:2moons} shows the positions of the Moon corresponding to the sunspot contact times. There are two points of intersection of the circles of the Moon, but only one point is located on the solar disk. This point corresponds to a heliographic latitude of $b=8.0^{\circ}$ and a Carrington longitude of $L=5.9^{\circ}$.

\begin{table}
\caption{Estimated contact times of the solar eclipse on 1816 November~19.}
\centering
\begin{tabular}{lllll}
\hline
   Event & Recorded time & UTC time \\
\hline
    Start of eclipse&  \phantom{0}8:27:12 &  \phantom{0}8:04:15\\
    Spot contact 1	&  \phantom{0}9:21:27 &  \phantom{0}8:58:39\\
    Spot contact 2	&            10:30:34 &            10:07:45\\
    End of eclipse	&            10:46:28 &            10:23:39\\
\hline
\end{tabular}
\label{tab:eclipse1816}
\end{table}

\begin{figure}
\centering
\includegraphics[width=0.8\columnwidth]{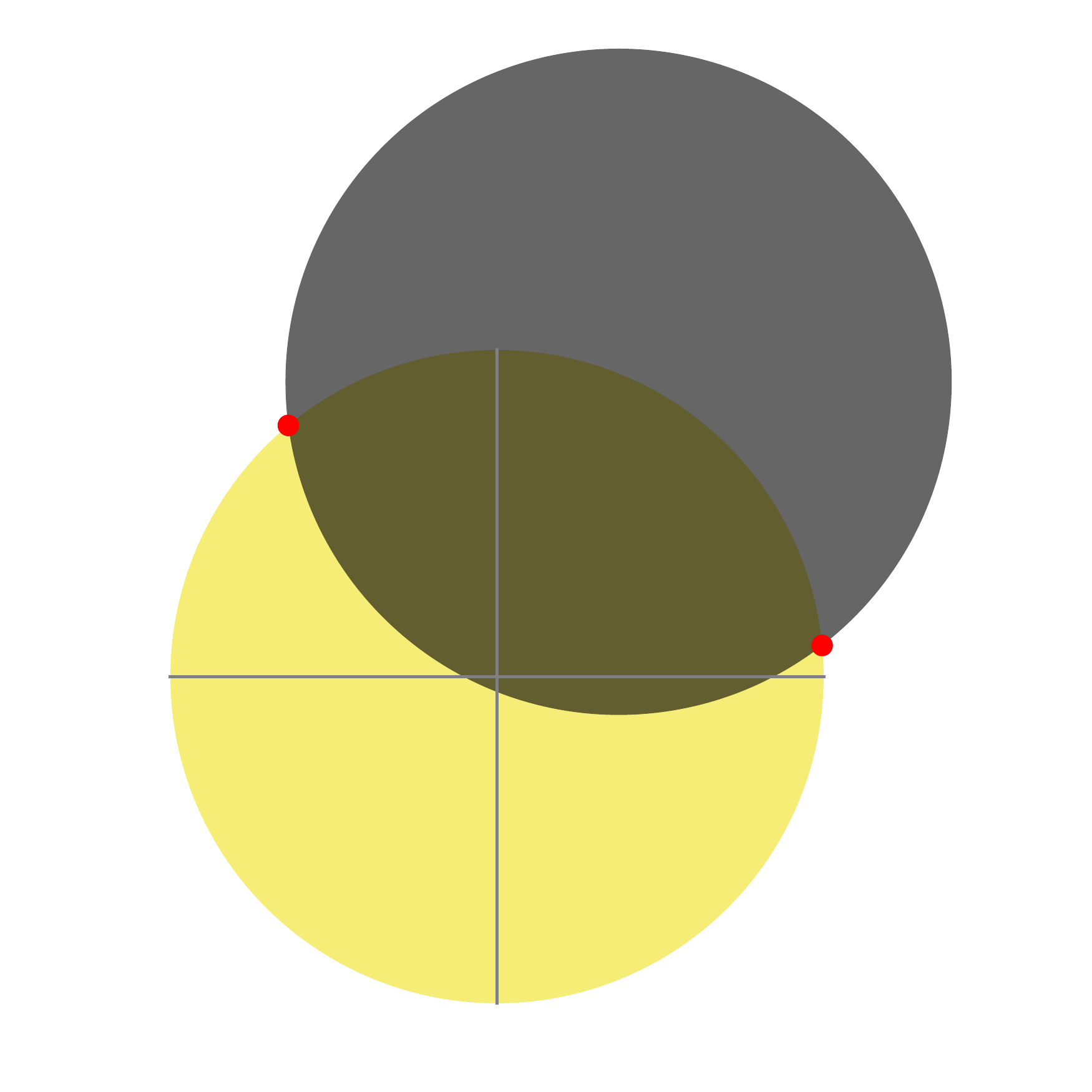}
    \caption{Apparent positions of the Sun and the Moon reconstructed with {\sc astropy} using the times of Table~\ref{tab:horns}. The red dots in the intersections of the solar disk with the lunar disk are the horns of the solar disk, labeled as ${\rm C}'$  and ${\rm C}''$ in Table~\ref{tab:horns}.}
    \label{fig:horns}
\end{figure}

During the solar eclipse described above, Flaugergues also measured the positions of the horns of the partially eclipsed solar disk using the rhomboid scheme (see Table~\ref{tab:horns} for a sample record). Fig.~\ref{fig:horns} shows the apparent positions of the Sun and the Moon reconstructed with {\sc astropy} using the times of Table~\ref{tab:horns}. Then we reconstruct the heliographic coordinates of the two horns ${\rm C}'$ and ${\rm C}''$, using the method described in Section~\ref{sec:rhomb}.

\begin{table}
\caption{Observation of the solar eclipse on 1816 November~19. The symbols ${\rm C}'$  and ${\rm C}''$ correspond to the two horns of the partially eclipsed solar disk.}
\centering
\begin{tabular}{llll}
\hline
Object & Event & Recorded time \\
\hline
Sun &    in  & 21:05:22 \\
${\rm C}'$ &    in & 21:05:35 \\
${\rm C}'$ &    out & 21:06:44 \\
${\rm C}''$ &   in & 21:07:49 \\
${\rm C}''$ & out & 21:08:16 \\ 
Sun &    out & 21:09:16 \\
\hline
\end{tabular}
\label{tab:horns}
\end{table}

\begin{figure}
\centering
\includegraphics[width=0.9\columnwidth]{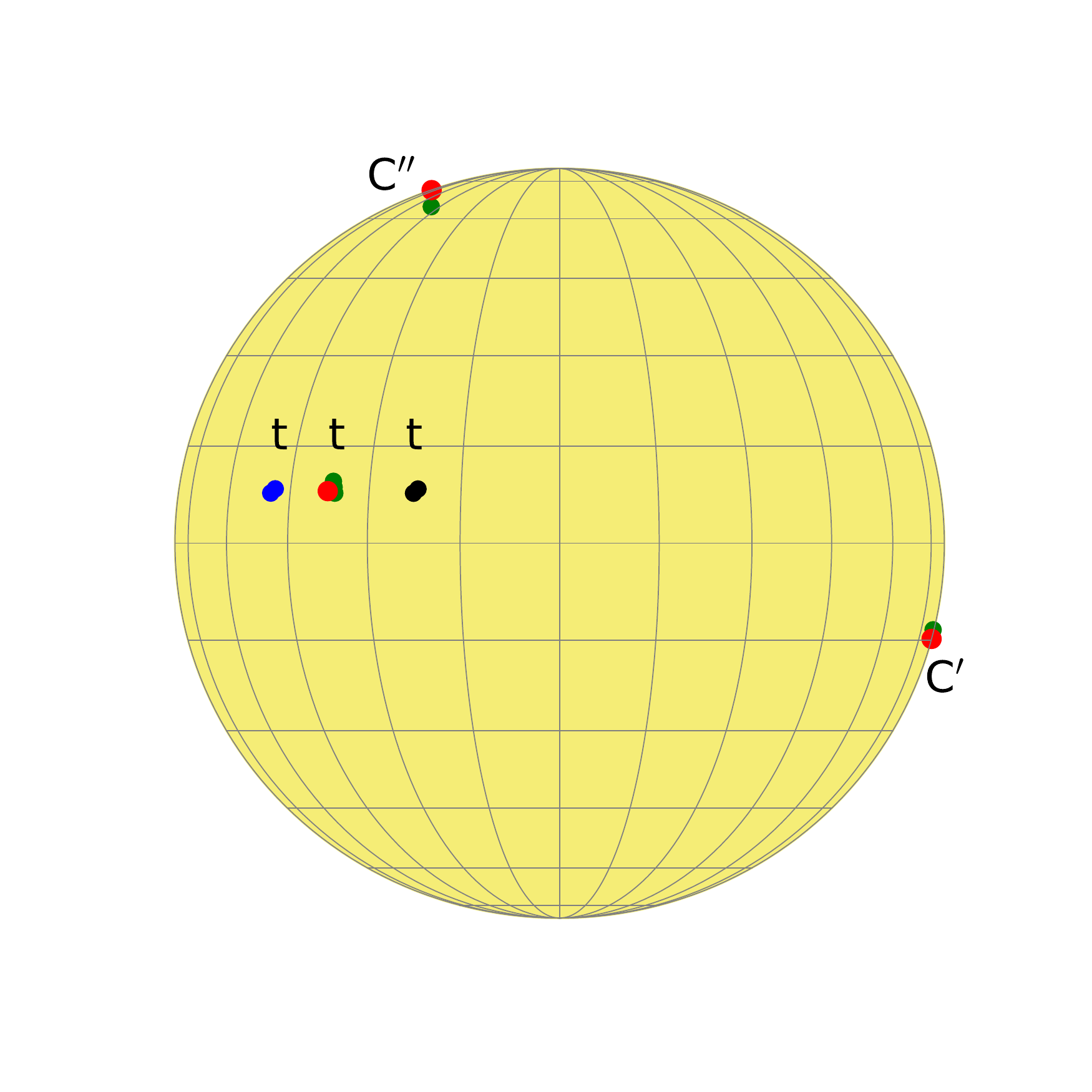}
    \caption{Red dots show the solar disk horns (${\rm C}'$  and ${\rm C}''$) and the sunspot (t) positions obtained by reconstructing the solar eclipse of 1816 November~19 using {\sc astropy}. The green dots show the corresponding positions reconstructed from two transit time measurements. The blue and black points are sunspot positions reconstructed on the previous and the next day.}
    \label{fig:eclipse_spots}
\end{figure}

Now we compare the results. Fig.~\ref{fig:eclipse_spots} shows the solar disk horns and the sunspot position obtained with {\sc astropy} (red dots) in comparison to the positions reconstructed from transit time records (green dots).
We observe that both methods give approximately the same coordinates
for the sunspot and both solar disk horns (see Table~\ref{tab:ecl_spots}).
We consider this consistency as an independent validation of the correctness of the coordinate reconstruction procedure.

\begin{table}
\caption{Coordinates of the sunspot observed on 1816 November~19
reconstructed from eclipse measurements and from two records made with the rhombus.}
\centering
\begin{tabular}{llll}
\hline
Method & Latitude & Longitude \\
\hline
Eclipse &   $8.0\degr$ & $5.94\degr$ \\
Rhombus &   $7.52\degr$ & $6.44\degr$ \\
Rhombus &    $8.43\degr$ & $6.17\degr$ \\
\hline
\end{tabular}
\label{tab:ecl_spots}
\end{table}

A more detailed investigation of the accuracy is provided in the Appendix based on the Mercury transit of 1799.

\begin{figure}
\centering
\includegraphics[width=0.9\columnwidth]{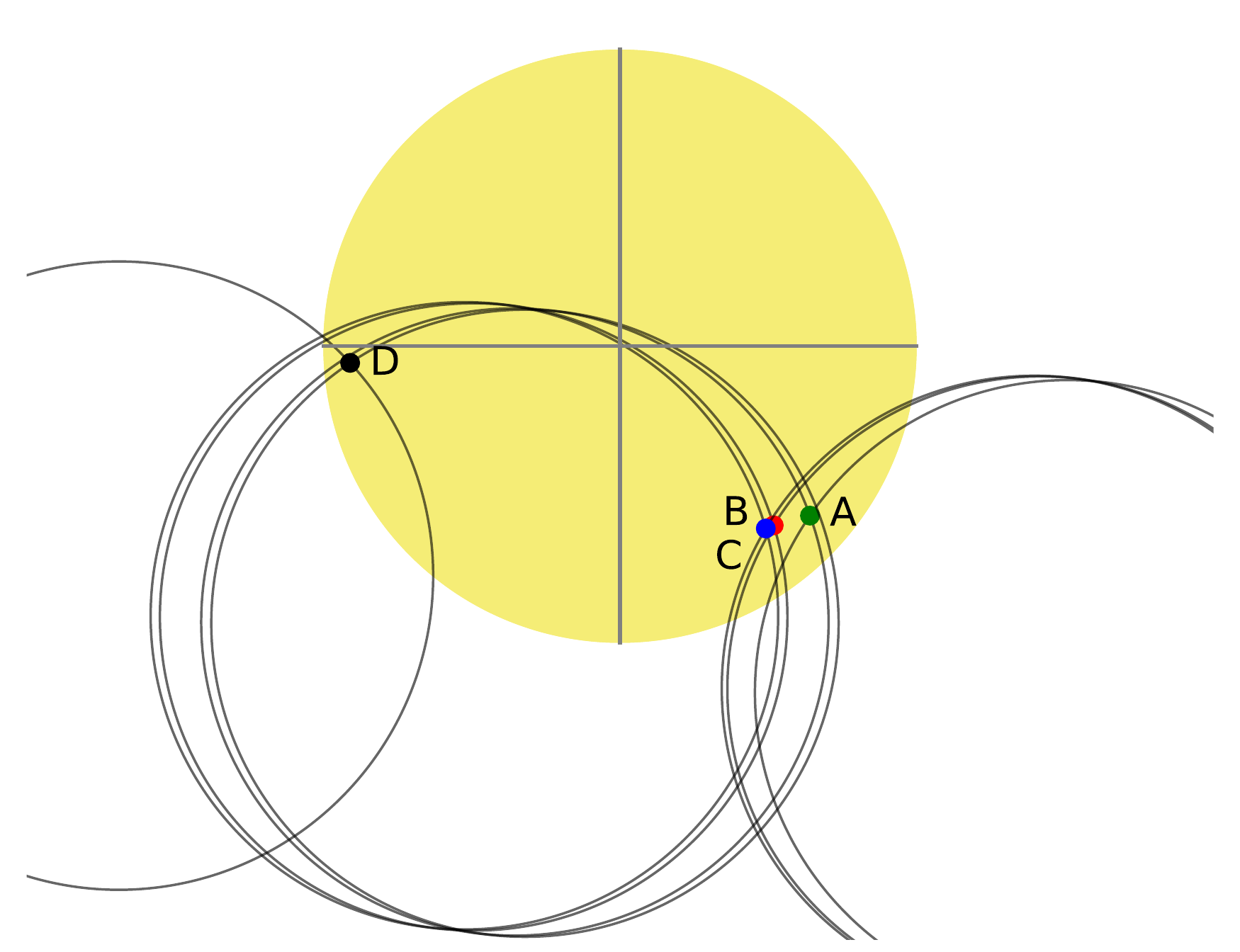}
\includegraphics[width=0.9\columnwidth]{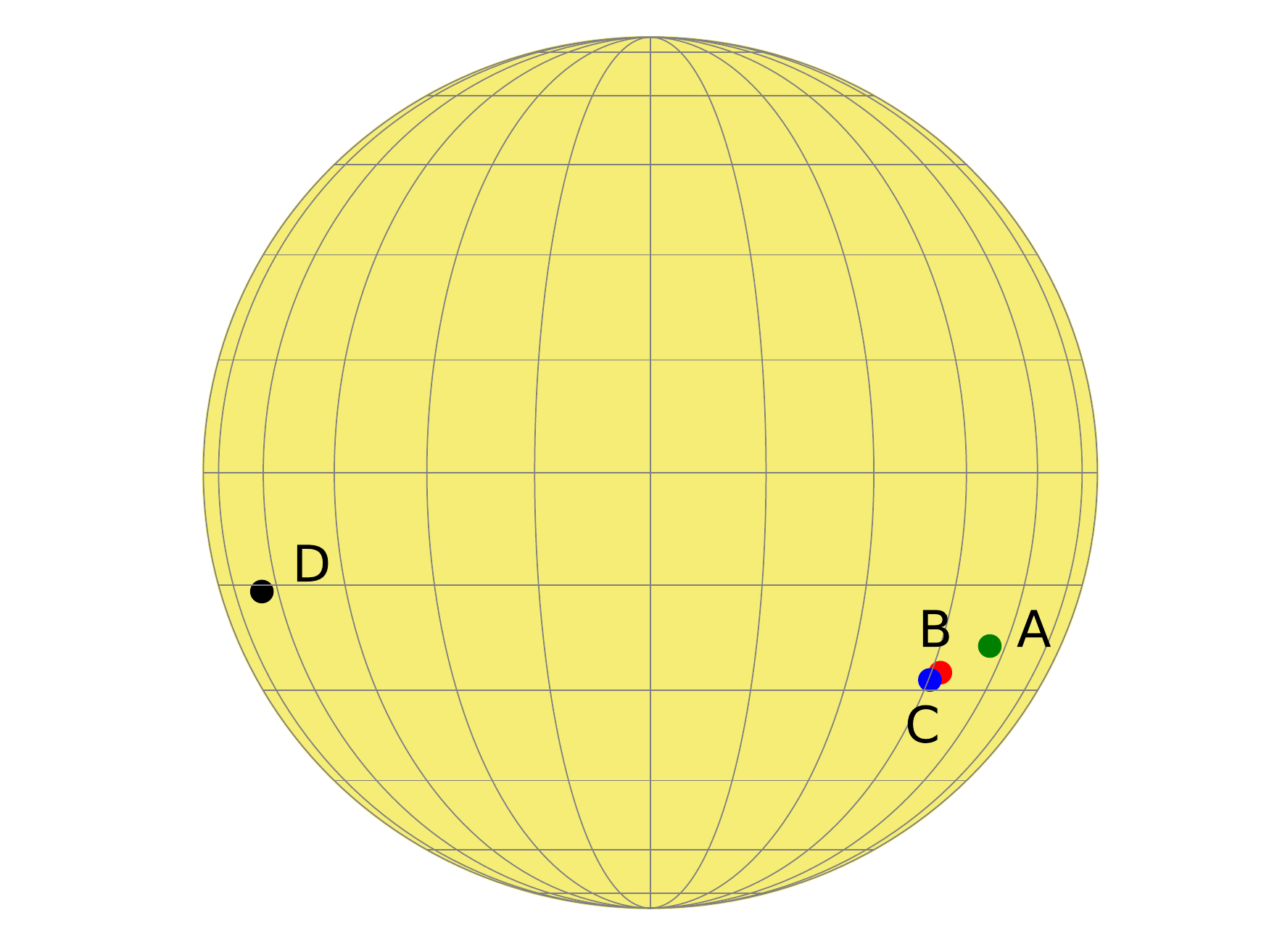}
\caption{Solar eclipse of 1788 June~4 reconstructed using {\sc astropy}. Top panel: black circles show positions of the Moon at the moments, when the sunspots contact the disk of the Moon (Table~\ref{tab:eclipse1788}). The corresponding intersections of the disks of the Moon are marked with coloured dots. The vertical points to the celestial north. The Moon moves from right to left with time. Bottom panel: sunspot positions reconstructed in heliographic coordinates.}
    \label{fig:eclipse_1788}
\end{figure}

Using the same approach, we reconstruct sunspot positions measured during the solar eclipse on 1788 June~4. They are the only positions we could derive from Flaugergues's observations before 1795. Table~\ref{tab:eclipse1788} contains the original records of the eclipse and four sunspots as well as the times corrected to UTC. The difference between the estimated and the observed solar eclipse duration is 26~s. Fig.~\ref{fig:eclipse_1788} shows the apparent positions of the Sun and the Moon at the moments, when the sunspots contact the disk of the Moon. The corresponding intersections of the lunar disks give the sunspot positions. The reconstructed sunspot positions in heliographic coordinates are shown in Fig.~\ref{fig:eclipse_1788} as well.

\begin{table}
\caption{Estimated contact times of the solar eclipse on 1788~June~4.}
\centering
\begin{tabular}{lllll}
\hline
   Event & Recorded time & UTC time \\
\hline
    Start of eclipse	& 7:26:36 & 7:05:30 \\
    Spot A contact 1 &  7:31:40 & 7:10:36 \\
    Spot B contact 1 &  7:34:32 &  7:13:28 \\
    Spot C contact 1 & 7:35:07 & 7:14:04 \\
    Spot D contact 1 & 8:30:24 & 8:09:32 \\
    Spot A contact 2 & 8:31:32 & 8:10:40 \\
    Spot B contact 2 & 8:36:12 &  8:15:21 \\
    Spot C contact 2 & 8:37:15 &  8:16:24 \\
    Spot D contact 2 & 9:17:32 &  8:56:49 \\
    End of eclipse   & 9:25:43 & 9:05:02 \\
\hline
\end{tabular}
\label{tab:eclipse1788}
\end{table}

\begin{figure}
\centering
\includegraphics[width=0.95\columnwidth]{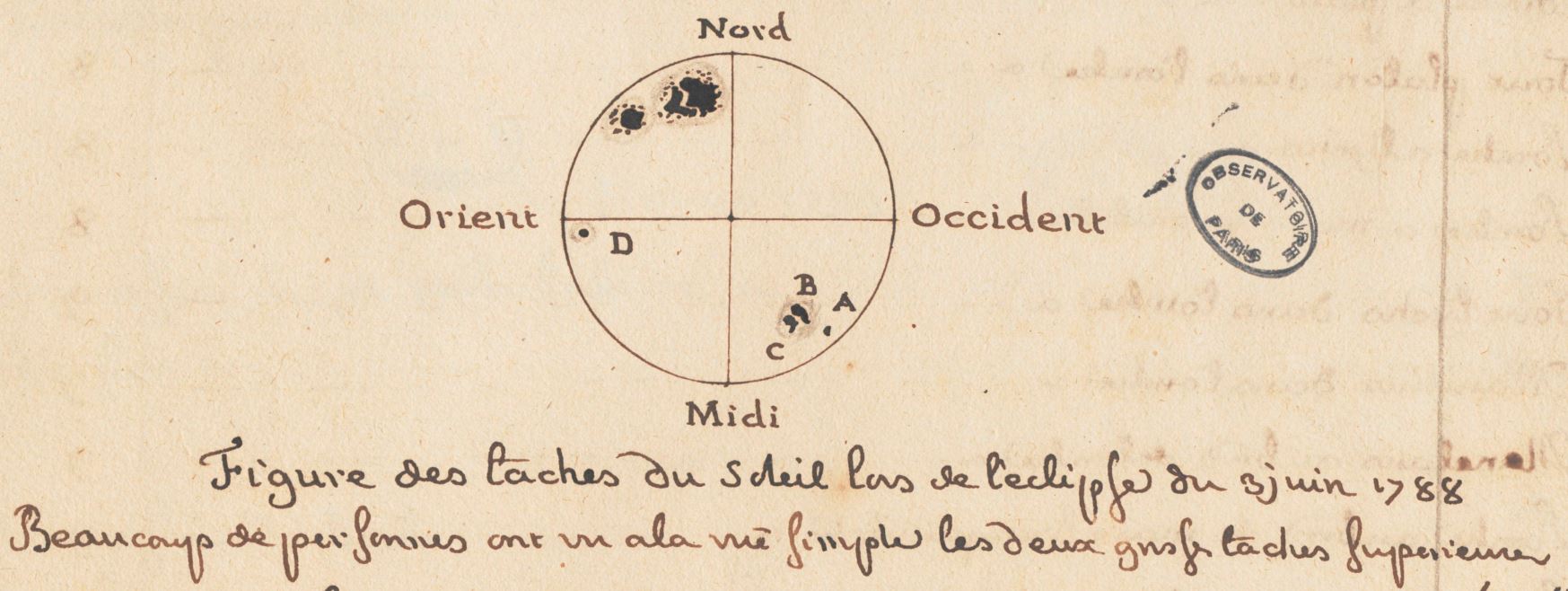}
    \caption{Original drawing with sunspot positions on 1788~June~4. The bottom line can be translated as
    ``many people on the street saw the two large upper spots''.}
    \label{fig:eclipse_1788_original}
\end{figure}

Interestingly, we find a drawing of the solar disk with sunspot positions made by Flaugergues on 1788 June~4. The original drawing (Fig.~\ref{fig:eclipse_1788_original}) looks quite close to the reconstructed positions (Fig.~\ref{fig:eclipse_1788}), however, the drawing additionally contains two unusually large spots in the top-left quadrant, which were not eclipsed by the Moon. These must indeed be spots, since we find a sentence below the drawing which can be translated as ``many people on the street saw the two large upper spots''. Note that according to, e.g., \cite{Keller} or \cite{Schaefer1993}, the limiting sunspot diameter visible to normal eye-sight people is about $40\arcsec$, but this group must have been considerably larger to be visible to casual witnesses.

It should be noted that the whole-disk drawings similar to Fig.~\ref{fig:eclipse_1788_original} are
very rare in Flaugergues books. We found only 20~drawings containing the whole disk or parts of it
which can be complemented to a full disk.

\begin{figure*}
\centering
\includegraphics[width=1\textwidth]{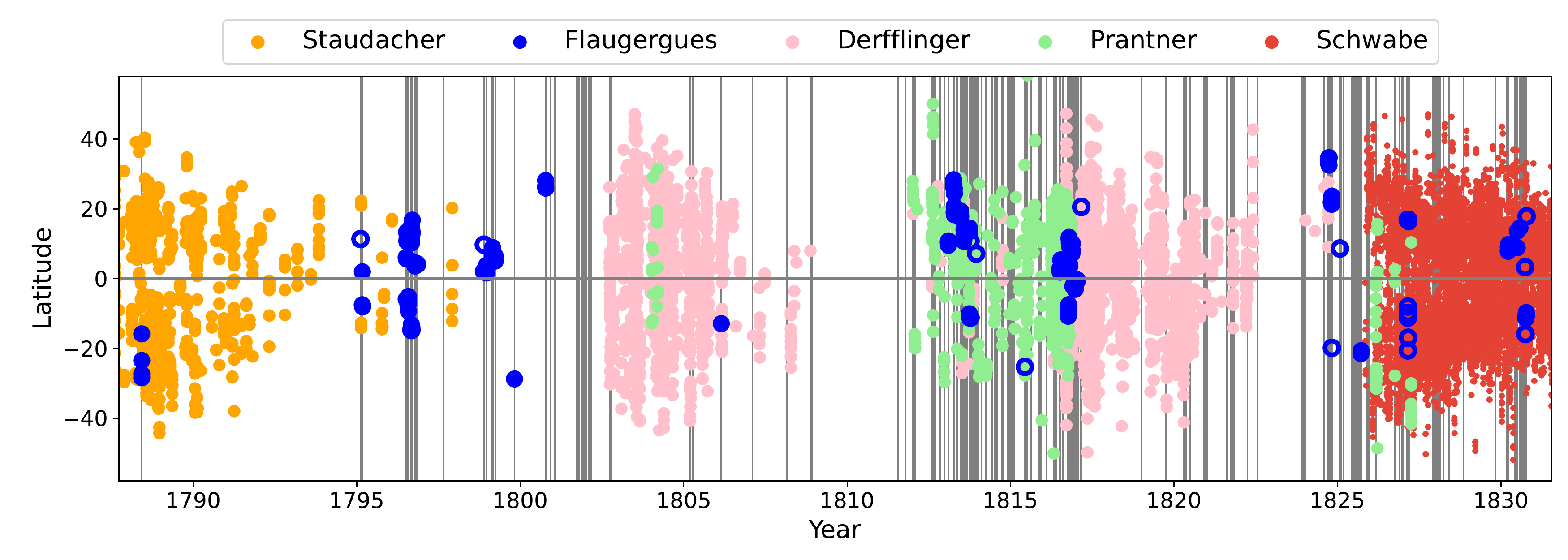}
    \caption{Time--latitude distribution of the sunspots positions obtained from Flaugergues's observations (in blue). The filled blue circles indicate reliable sunspot positions, whereas open blue circles indicate uncertain positions. Days with transit records are marked by gray vertical lines. For comparison, we also plot sunspot positions from
    previously reconstructed data. The orange dots show sunspot distribution from the Staudacher and Armagh catalogues \citep{arlt2009,arlt2009armagh},
    the pink dots show sunspot distribution from the Derfflinger catalogue \citep{hayakawa_ea2020},
    the green dots correspond to the sunspots reconstructed from Prantner's drawings \citep{Hayakawa_2021}, and
    the red dots correspond to the sunspots reconstructed from Schwabe's drawings \citep{arlt_ea2013}. 
    }
    \label{fig:timelat}
\end{figure*}

\section{Results and Discussion}

In total, we were able to reconstruct 527 sunspot positions for 154~days out of 156~days for which appropriate transits were recorded. Together with single-transit measurements (which do not allow a direct coordinate reconstruction), there are 463~days of documented sunspot observations. 
We combined repeated observations of the same spots on the same day into averaged positions
and  resulted in 196~positions.
Then we discriminated the positions subjectively into
reliable and less reliable ones, based on the completeness of information, the additional textual information given, and our general experience with the observations. There is no mathematical distinction between the two classes of positions.
Fig.~\ref{fig:timelat} shows the time--latitude distribution obtained.

In the process of reconstructing the coordinates, we found a number of errors in writing among
the original records. We assumed the presence of errors when a group of several records had one sunspot measurement with heliographic coordinates not matching the other records in this group. Another case of suspected errors is when a sunspot on a particular day can not be matched with any sunspot before or after that day.
When we detected such potential errors, we tried to correct them by assuming that either two lines in the record are interchanged
or one minute should be added or subtracted from the recorded time (this modifies a single digit in the record).
We avoided corrections that require changing two or more digits in the record because the probability of such error
is much lower. With these two options, we were able to bring the sunspot-outliers into positions consistent with the other records in most cases.
In the output files, we provide both the original times and our modifications. 
The reconstructed coordinates presented here take the modifications into account.

Fig.~\ref{fig:timelat} admits the identification of four solar cycles. Before 1800 we observe the end of cycle~4. Sunspots observed at high latitudes ($\pm 30^{\circ}$) near 1800 indicate the beginning of cycle~5 which lasted until the end of 1810 \citep{hathaway2015}. We then find an increased density of observations and several sunspots at high latitudes indicating the beginning of the cycle~6. The low number of records in 1823--1824 very likely corresponds to the solar minimum followed by high latitudes of spots of cycle~7 in 1825. 

In may be noted that in cycle~6 the number of sunspots in the 
northern hemisphere is larger than the number of sunspots in the
southern hemisphere, while during the Maunder Minimum, sunspot
activity prevailed in the southern hemisphere \citep{Ribes}.
This result may be a small-number effect and requires confirmation.
It is clear from his books that Flaugergues did not record all the 
sunspots he observed. Also Hallaschka appears to have recorded a 
slightly higher number of spots in the northern hemisphere 
than in the southern one, according to the results by
\citet{carrasco_ea2018}, who placed reasonable best-guess equators 
into the drawings, whose orientations are difficult to assess by
objective means. Similarly to Flaugergues's observations, 
the dominance of northern-hemisphere spots in Derfflinger's data 
\citep{hayakawa_ea2020} is noticeable only in the years 1812--1814 and 
is less obvious in Prantner's observations of the beginning of cycle~6 
\citep{Hayakawa_2021}.

Unfortunately, there is no spot measurement in 1793 or 1794 when
spot latitudes may shed light on the suspected short cycle between
cycles~4 and~5 (lost cycle), beyond the spot latitudes scrutinized by \citet{usoskin_ea2009}. The latitudes observed particularly in 1796 support the relatively wide butterfly diagram derived from Staudacher and the Armagh observations \citep{arlt2009,arlt2009armagh} which follow very low latitudes in 1792 observed by Staudacher. 

The high latitudes observed by Flaugergues near 1800 indicate that the lost cycle did not last longer than the year 1800. We do see a remarkable decrease in absolute latitudes from 1796 to 1799 in the data by Flaugergues, supporting the quick evolution of a short cycle before 1800.

We found several examples of a sunspot (or sunspot group) living for more than one solar rotation period. The first example starts on 1796 July~5 with a spot denoted originally by the letter~A and continues on 1796 July~24 with a spot denoted by~H (Table~\ref{tab:rot1796}).

\begin{table}
\caption{Example 1 of a sunspot observed for more than one solar rotation period.
CMD is the heliocentric central-meridian distance.}
\centering
\begin{tabular}{llccr}
\hline
   Date & Spot & Latitude & Longitude & CMD \\
\hline
    1796 Jul 5	& A	& $-5.92\degr$	&   $29.47\degr$	&	$52.85\degr$ \\
    1796 Jul 24	& H	& $-5.95\degr$	&	$33.09\degr$	&	$-53.39\degr$ \\
    1796 Jul 25	& H	& $-6.67\degr$	&	$31.93\degr$	&	$-40.97\degr$ \\
    1796 Jul 26	& H	& $-7.59\degr$	&	$32.03\degr$	&	$-27.75\degr$ \\
    1796 Jul 28	& t	& $-7.35\degr$	&	$30.56\degr$	&	$-2.62\degr$ \\
    1796 Jul 29	& H	& $-7.22\degr$	&	$31.59\degr$	&	$12.18\degr$ \\
    1796 Jul 30	& H	& $-6.88\degr$	&	$32.41\degr$	&	$25.76\degr$ \\
    1796 Jul 31	& H	& $-9.32\degr$	&	$31.95\degr$	&	$38.07\degr$  \\
 \\
\hline
\end{tabular}
\label{tab:rot1796}
\end{table}

The second example starts on 1799 February~24 and consists of the two spots given in Table~\ref{tab:rot1799}. About one rotation later, on 1799 March~27 and~28, we find a sunspot with approximately the same coordinates as the trailing spot of the earlier group.

\begin{table}
\caption{Example 2 of sunspots observed for more than one solar rotation period. CMD is the heliocentric central-meridian distance.}
\centering
\begin{tabular}{llccr}
\hline
   Date & Spot & Latitude & Longitude & CMD \\
\hline
    1799 Feb 24	& ${\rm t}_1$	& $7.30\degr$	& $214.48\degr$ & 	$-0.85\degr$ \\
    1799 Feb 24	& ${\rm t}_2$	& $5.63\degr$	& $202.51\degr$ & 	$-12.82\degr$ \\
    1799 Feb 26	& ${\rm t}_1$	& $8.87\degr$	& $215.59\degr$ & 	$26.69\degr$ \\
    1799 Feb 26	& ${\rm t}_2$	& $6.54\degr$	& $203.40\degr$ & 	$14.51\degr$ \\
    1799 Mar 27	& t	            & $6.52\degr$	& $204.52\degr$ & 	$39.77\degr$ \\
    1799 Mar 28	& t	            & $5.02\degr$	& $204.67\degr$ & 	$54.93\degr$ \\
\hline
\end{tabular}
\label{tab:rot1799}
\end{table}

One more example is the sunspot observed on 1816 November~26 and November~27, and then on 1816 December~17 and several days thereafter (see Table~\ref{tab:rot1816}). In the following days, this sunspot drifts rather unusually in longitude, remaining at about the same latitude, which may be due to the low accuracy of sunspot measurements near the solar limb.

\begin{table}
\caption{Example 3 of a sunspot observed for more than one solar rotation period. CMD is the heliocentric central-meridian distance.}
\centering
\begin{tabular}{llcrr}
\hline
    Date & Spot & Latitude & Longitude & CMD \\
\hline
    1816 Nov 26	& t	& $-2.15\degr$	& $6.40\degr$	& $57.76\degr$ \\
    1816 Nov 27	& t	& $-2.13\degr$	& $6.02\degr$	& $70.30\degr$ \\ 
    1816 Dec 17	& t	& $-1.48\degr$	& $5.37\degr$	& $-26.55\degr$ \\
    1816 Dec 18	& t	& $-0.45\degr$	& $7.09\degr$	& $-11.80\degr$ \\ 
    1816 Dec 20	& t	& $-2.82\degr$	& $9.05\degr$	& $16.72\degr$ \\
    1816 Dec 21	& t	& $-2.42\degr$	& $10.38\degr$	& $30.71\degr$ \\
    1816 Dec 23	& t	& $-3.07\degr$	& $13.58\degr$	& $60.25\degr$ \\
    1816 Dec 24	& t	& $-2.72\degr$	& $22.58\degr$	& $83.01\degr$ \\
\hline
\end{tabular}
\label{tab:rot1816}
\end{table}

While we have not touched the issue of sunspot numbers and group sunspot numbers from the observations by Flaugergues in the present paper, the positions derived may be useful in discriminating sunspot groups through the angular distance of spots in the visible solar hemisphere. A potential revision of the sunspot numbers from Flaugergues is suggested for the future.
At this stage, we note that the reconstructed sunspot positions underestimate the solar activity
since Flaugergues did not measure contact  times for all sunspots he observed. We expect that 
detailed translations of the textual notes, which we left out of the scope of this study, may help
to clarify the situation at least in some cases. In particular, some of the textual notes indicate spotless days,
others describe sunspot configurations on the disk.
We hope that this article will rouse interest in further investigations of these manuscripts.

We provide the numerical observational records and the reconstructed coordinates in the GitHub repository \href{https://github.com/observethesun/Flaugergues.git}{github.com/observethesun/Flaugergues}.

\section*{Acknowledgements}
We would like to thank the Biblioth\`eque de l'Observatoire de Paris for the inspection and
digitization of the observing books by Flaugergues.
The authors are grateful to Deutsche Forschungsgemeinschaft for their grant Ar/355-12 and to
DAAD Forschungsstipendium allowing for the digitization of the manuscripts and the visits of EI to Potsdam.
EI acknowledges the support of RSF grant 21-72-20067 for the creation of the database.
The authors are grateful for the valuable comments by the reviewer of the paper.

\section*{Data Availability}
The data underlying this article are available in GitHub at \href{https://github.com/observethesun/Flaugergues.git}{github.com/observethesun/Flaugergues}.

\bibliographystyle{mnras}
\bibliography{cite.bib} 

\begin{thebibliography}{}
\makeatletter
\relax
\def\mn@urlcharsother{\let\do\@makeother \do\$\do\&\do\#\do\^\do\_\do\%\do\~}
\def\mn@doi{\begingroup\mn@urlcharsother \@ifnextchar [ {\mn@doi@}
  {\mn@doi@[]}}
\def\mn@doi@[#1]#2{\def\@tempa{#1}\ifx\@tempa\@empty \href
  {http://dx.doi.org/#2} {doi:#2}\else \href {http://dx.doi.org/#2} {#1}\fi
  \endgroup}
\def\mn@eprint#1#2{\mn@eprint@#1:#2::\@nil}
\def\mn@eprint@arXiv#1{\href {http://arxiv.org/abs/#1} {{\tt arXiv:#1}}}
\def\mn@eprint@dblp#1{\href {http://dblp.uni-trier.de/rec/bibtex/#1.xml}
  {dblp:#1}}
\def\mn@eprint@#1:#2:#3:#4\@nil{\def\@tempa {#1}\def\@tempb {#2}\def\@tempc
  {#3}\ifx \@tempc \@empty \let \@tempc \@tempb \let \@tempb \@tempa \fi \ifx
  \@tempb \@empty \def\@tempb {arXiv}\fi \@ifundefined
  {mn@eprint@\@tempb}{\@tempb:\@tempc}{\expandafter \expandafter \csname
  mn@eprint@\@tempb\endcsname \expandafter{\@tempc}}}

\bibitem[\protect\citeauthoryear{{Arlt}}{{Arlt}}{2009a}]{arlt2009}
{Arlt} R.,  2009a, \mn@doi [Solar Phys.] {10.1007/s11207-008-9306-5}, \href
  {http://adsabs.harvard.edu/abs/2009SoPh..255..143A} {255, 143}

\bibitem[\protect\citeauthoryear{{Arlt}}{{Arlt}}{2009b}]{arlt2009armagh}
{Arlt} R.,  2009b, \mn@doi [Astronomische Nachrichten]
  {10.1002/asna.200911195}, \href
  {http://adsabs.harvard.edu/abs/2009AN....330..311A} {330, 311}

\bibitem[\protect\citeauthoryear{{Arlt}, {Leussu}, {Giese}, {Mursula}  \&
  {Usoskin}}{{Arlt} et~al.}{2013}]{arlt_ea2013}
{Arlt} R.,  {Leussu} R.,  {Giese} N.,  {Mursula} K.,   {Usoskin} I.~G.,  2013,
  \mn@doi [\mnras] {10.1093/mnras/stt961}, \href
  {https://ui.adsabs.harvard.edu/abs/2013MNRAS.433.3165A} {433, 3165}

\bibitem[\protect\citeauthoryear{{Astropy Collaboration} et~al.,}{{Astropy
  Collaboration} et~al.}{2013}]{astropy:2013}
{Astropy Collaboration} et~al., 2013, \mn@doi [\aap]
  {10.1051/0004-6361/201322068}, \href
  {http://adsabs.harvard.edu/abs/2013A%26A...558A..33A} {558, A33}

\bibitem[\protect\citeauthoryear{{Astropy Collaboration} et~al.,}{{Astropy
  Collaboration} et~al.}{2018}]{astropy:2018}
{Astropy Collaboration} et~al., 2018, \mn@doi [\aj] {10.3847/1538-3881/aabc4f},
  \href {https://ui.adsabs.harvard.edu/abs/2018AJ....156..123A} {156, 123}

\bibitem[\protect\citeauthoryear{{Astropy Collaboration} et~al.,}{{Astropy
  Collaboration} et~al.}{2022}]{astropy:2022}
{Astropy Collaboration} et~al., 2022, \mn@doi [\apj]
  {10.3847/1538-4357/ac7c74}, \href
  {https://ui.adsabs.harvard.edu/abs/2022ApJ...935..167A} {935, 167}

\bibitem[\protect\citeauthoryear{{Asvestari}, {Usoskin}, {Kovaltsov}, {Owens},
  {Krivova}, {Rubinetti}  \& {Taricco}}{{Asvestari}
  et~al.}{2017}]{asvestari_ea2017}
{Asvestari} E.,  {Usoskin} I.~G.,  {Kovaltsov} G.~A.,  {Owens} M.~J.,
  {Krivova} N.~A.,  {Rubinetti} S.,   {Taricco} C.,  2017, \mn@doi [\mnras]
  {10.1093/mnras/stx190}, \href
  {https://ui.adsabs.harvard.edu/abs/2017MNRAS.467.1608A} {467, 1608}

\bibitem[\protect\citeauthoryear{{Berggren} et~al.,}{{Berggren}
  et~al.}{2009}]{berggren_ea2009}
{Berggren} A.~M.,  et~al., 2009, \mn@doi [\grl] {10.1029/2009GL038004}, \href
  {https://ui.adsabs.harvard.edu/abs/2009GeoRL..3611801B} {36, L11801}

\bibitem[\protect\citeauthoryear{{Brehm} et~al.,}{{Brehm}
  et~al.}{2021}]{Brehm2021}
{Brehm} N.,  et~al., 2021, \mn@doi [Nature Geoscience]
  {10.1038/s41561-020-00674-0}, \href
  {https://ui.adsabs.harvard.edu/abs/2021NatGe..14...10B} {14, 10}

\bibitem[\protect\citeauthoryear{{Cameron} \& {Sch{\"u}ssler}}{{Cameron} \&
  {Sch{\"u}ssler}}{2019}]{Cameron2019}
{Cameron} R.~H.,  {Sch{\"u}ssler} M.,  2019, \mn@doi [\aap]
  {10.1051/0004-6361/201935290}, \href
  {https://ui.adsabs.harvard.edu/abs/2019A&A...625A..28C} {625, A28}

\bibitem[\protect\citeauthoryear{{Carrasco}}{{Carrasco}}{2021}]{carrasco2021}
{Carrasco} V.~M.~S.,  2021, \mn@doi [\apj] {10.3847/1538-4357/ac24a5}, \href
  {https://ui.adsabs.harvard.edu/abs/2021ApJ...922...58C} {922, 58}

\bibitem[\protect\citeauthoryear{{Carrasco}, {Vaquero}, {Arlt}  \&
  {Gallego}}{{Carrasco} et~al.}{2018}]{carrasco_ea2018}
{Carrasco} V.~M.~S.,  {Vaquero} J.~M.,  {Arlt} R.,   {Gallego} M.~C.,  2018,
  \mn@doi [\solphys] {10.1007/s11207-018-1322-5}, \href
  {https://ui.adsabs.harvard.edu/abs/2018SoPh..293..102C} {293, 102}

\bibitem[\protect\citeauthoryear{{Carrasco}, {Hayakawa}, {Kuroyanagi},
  {Gallego}  \& {Vaquero}}{{Carrasco} et~al.}{2021}]{Carrasco_ea2021}
{Carrasco} V.~M.~S.,  {Hayakawa} H.,  {Kuroyanagi} C.,  {Gallego} M.~C.,
  {Vaquero} J.~M.,  2021, \mn@doi [\mnras] {10.1093/mnras/stab1155}, \href
  {https://ui.adsabs.harvard.edu/abs/2021MNRAS.504.5199C} {504, 5199}

\bibitem[\protect\citeauthoryear{{Carrasco}, {Llera}, {Aparicio}, {Gallego}  \&
  {Vaquero}}{{Carrasco} et~al.}{2022}]{carrasco_ea2022}
{Carrasco} V.~M.~S.,  {Llera} J.,  {Aparicio} A.~J.~P.,  {Gallego} M.~C.,
  {Vaquero} J.~M.,  2022, \mn@doi [\apj] {10.3847/1538-4357/ac7045}, \href
  {https://ui.adsabs.harvard.edu/abs/2022ApJ...933...26C} {933, 26}

\bibitem[\protect\citeauthoryear{{Connor}}{{Connor}}{1953}]{Lalande}
{Connor} E.,  1953, Leaflet of the Astronomical Society of the Pacific, \href
  {https://ui.adsabs.harvard.edu/abs/1953ASPL....6..330C} {6, 330}

\bibitem[\protect\citeauthoryear{Flaugergues}{Flaugergues}{1814}]{fla}
Flaugergues H.,  1813-1814, Annales de math\'ematiques pures et appliqu\'ees,
  4, 321

\bibitem[\protect\citeauthoryear{{Gao}}{{Gao}}{2017}]{gao2017}
{Gao} P.~X.,  2017, \mn@doi [\mnras] {10.1093/mnras/stx2206}, \href
  {https://ui.adsabs.harvard.edu/abs/2017MNRAS.472.2913G} {472, 2913}

\bibitem[\protect\citeauthoryear{{Hathaway}}{{Hathaway}}{2015}]{hathaway2015}
{Hathaway} D.~H.,  2015, \mn@doi [Living Reviews in Solar Physics]
  {10.1007/lrsp-2015-4}, \href
  {https://ui.adsabs.harvard.edu/abs/2015LRSP...12....4H} {12, 4}

\bibitem[\protect\citeauthoryear{{Hayakawa}, {Besser}, {Iju}, {Arlt}, {Uneme},
  {Imada}, {Bourdin}  \& {Kraml}}{{Hayakawa} et~al.}{2020}]{hayakawa_ea2020}
{Hayakawa} H.,  {Besser} B.~P.,  {Iju} T.,  {Arlt} R.,  {Uneme} S.,  {Imada}
  S.,  {Bourdin} P.-A.,   {Kraml} A.,  2020, \mn@doi [\apj]
  {10.3847/1538-4357/ab65c9}, \href
  {https://ui.adsabs.harvard.edu/abs/2020ApJ...890...98H} {890, 98}

\bibitem[\protect\citeauthoryear{{Hayakawa}, {Kuroyanagi}, {Carrasco}, {Uneme},
  {Besser}, {S{\^o}ma}  \& {Imada}}{{Hayakawa} et~al.}{2021a}]{Hayakawa2021}
{Hayakawa} H.,  {Kuroyanagi} C.,  {Carrasco} V. M.~S.,  {Uneme} S.,  {Besser}
  B.~P.,  {S{\^o}ma} M.,   {Imada} S.,  2021a, \mn@doi [\apj]
  {10.3847/1538-4357/abd949}, \href
  {https://ui.adsabs.harvard.edu/abs/2021ApJ...909..166H} {909, 166}

\bibitem[\protect\citeauthoryear{{Hayakawa}, {Uneme}, {Besser}, {Iju}  \&
  {Imada}}{{Hayakawa} et~al.}{2021b}]{Hayakawa_2021}
{Hayakawa} H.,  {Uneme} S.,  {Besser} B.~P.,  {Iju} T.,   {Imada} S.,  2021b,
  \mn@doi [\apj] {10.3847/1538-4357/abee1b}, \href
  {https://ui.adsabs.harvard.edu/abs/2021ApJ...919....1H} {919, 1}

\bibitem[\protect\citeauthoryear{{Hoyt} \& {Schatten}}{{Hoyt} \&
  {Schatten}}{1996}]{Hoyt1996}
{Hoyt} D.~V.,  {Schatten} K.~H.,  1996, \mn@doi [\solphys]
  {10.1007/BF00149097}, \href
  {https://ui.adsabs.harvard.edu/abs/1996SoPh..165..181H} {165, 181}

\bibitem[\protect\citeauthoryear{{Karoff}, {J{\o}rgensen}, {Senthamizh Pavai}
  \& {Arlt}}{{Karoff} et~al.}{2019}]{karoff_ea2019}
{Karoff} C.,  {J{\o}rgensen} C.~S.,  {Senthamizh Pavai} V.,   {Arlt} R.,  2019,
  \mn@doi [\solphys] {10.1007/s11207-019-1466-y}, \href
  {https://ui.adsabs.harvard.edu/abs/2019SoPh..294...78K} {294, 78}

\bibitem[\protect\citeauthoryear{{Keller} \& {Friedli}}{{Keller} \&
  {Friedli}}{1992}]{Keller}
{Keller} H.~U.,  {Friedli} T.~K.,  1992, \qjras, \href
  {https://ui.adsabs.harvard.edu/abs/1992QJRAS..33...83K} {33, 83}

\bibitem[\protect\citeauthoryear{{Miyahara}, {Masuda}, {Muraki}, {Furuzawa},
  {Menjo}  \& {Nakamura}}{{Miyahara} et~al.}{2004}]{miyahara_ea2004}
{Miyahara} H.,  {Masuda} K.,  {Muraki} Y.,  {Furuzawa} H.,  {Menjo} H.,
  {Nakamura} T.,  2004, \mn@doi [\solphys] {10.1007/s11207-005-6501-5}, \href
  {https://ui.adsabs.harvard.edu/abs/2004SoPh..224..317M} {224, 317}

\bibitem[\protect\citeauthoryear{{Nagovitsyn}}{{Nagovitsyn}}{2007}]{nagovitsyn2007}
{Nagovitsyn} Y.~A.,  2007, \mn@doi [Astronomy Letters]
  {10.1134/S1063773707050076}, \href
  {https://ui.adsabs.harvard.edu/abs/2007AstL...33..340N} {33, 340}

\bibitem[\protect\citeauthoryear{{Nielsen} \& {Kjeldsen}}{{Nielsen} \&
  {Kjeldsen}}{2011}]{Nielsen}
{Nielsen} M.~L.,  {Kjeldsen} H.,  2011, \mn@doi [\solphys]
  {10.1007/s11207-011-9733-6}, \href
  {https://ui.adsabs.harvard.edu/abs/2011SoPh..270..385N} {270, 385}

\bibitem[\protect\citeauthoryear{{Ribes} \& {Nesme-Ribes}}{{Ribes} \&
  {Nesme-Ribes}}{1993}]{Ribes}
{Ribes} J.~C.,  {Nesme-Ribes} E.,  1993, \aap, \href
  {https://ui.adsabs.harvard.edu/abs/1993A&A...276..549R} {276, 549}

\bibitem[\protect\citeauthoryear{{Schaefer}}{{Schaefer}}{1993}]{Schaefer1993}
{Schaefer} B.~E.,  1993, \mn@doi [\apj] {10.1086/172895}, \href
  {https://ui.adsabs.harvard.edu/abs/1993ApJ...411..909S} {411, 909}

\bibitem[\protect\citeauthoryear{{Silverman} \& {Hayakawa}}{{Silverman} \&
  {Hayakawa}}{2021}]{Silverman2021}
{Silverman} S.~M.,  {Hayakawa} H.,  2021, \mn@doi [Journal of Space Weather and
  Space Climate] {10.1051/swsc/202008210.48550/arXiv.2012.13713}, \href
  {https://ui.adsabs.harvard.edu/abs/2021JSWSC..11...17S} {11, 17}

\bibitem[\protect\citeauthoryear{{Stephenson}, {Morrison}  \&
  {Hohenkerk}}{{Stephenson} et~al.}{2016}]{Stephenson}
{Stephenson} F.~R.,  {Morrison} L.~V.,   {Hohenkerk} C.~Y.,  2016, \mn@doi
  [Proceedings of the Royal Society of London Series A]
  {10.1098/rspa.2016.0404}, \href
  {https://ui.adsabs.harvard.edu/abs/2016RSPSA.47260404S} {472, 20160404}

\bibitem[\protect\citeauthoryear{{Svalgaard} \& {Schatten}}{{Svalgaard} \&
  {Schatten}}{2016}]{Svalgaard2016}
{Svalgaard} L.,  {Schatten} K.~H.,  2016, \mn@doi [\solphys]
  {10.1007/s11207-015-0815-810.48550/arXiv.1506.00755}, \href
  {https://ui.adsabs.harvard.edu/abs/2016SoPh..291.2653S} {291, 2653}

\bibitem[\protect\citeauthoryear{{Usoskin}, {Mursula}  \&
  {Kovaltsov}}{{Usoskin} et~al.}{2003}]{Usoskin2003}
{Usoskin} I.~G.,  {Mursula} K.,   {Kovaltsov} G.~A.,  2003, \mn@doi [\solphys]
  {10.1023/B:SOLA.0000013029.99907.97}, \href
  {https://ui.adsabs.harvard.edu/abs/2003SoPh..218..295U} {218, 295}

\bibitem[\protect\citeauthoryear{{Usoskin}, {Mursula}, {Arlt}  \&
  {Kovaltsov}}{{Usoskin} et~al.}{2009}]{usoskin_ea2009}
{Usoskin} I.~G.,  {Mursula} K.,  {Arlt} R.,   {Kovaltsov} G.~A.,  2009, \mn@doi
  [\apjl] {10.1088/0004-637X/700/2/L154}, \href
  {https://ui.adsabs.harvard.edu/abs/2009ApJ...700L.154U} {700, L154}

\bibitem[\protect\citeauthoryear{{Usoskin} et~al.,}{{Usoskin}
  et~al.}{2015}]{usoskin_ea2015}
{Usoskin} I.~G.,  et~al., 2015, \mn@doi [\aap] {10.1051/0004-6361/201526652},
  \href {https://ui.adsabs.harvard.edu/abs/2015A&A...581A..95U} {581, A95}

\bibitem[\protect\citeauthoryear{{Usoskin}, {Solanki}, {Krivova}, {Hofer},
  {Kovaltsov}, {Wacker}, {Brehm}  \& {Kromer}}{{Usoskin}
  et~al.}{2021}]{Usoskin2021}
{Usoskin} I.~G.,  {Solanki} S.~K.,  {Krivova} N.~A.,  {Hofer} B.,  {Kovaltsov}
  G.~A.,  {Wacker} L.,  {Brehm} N.,   {Kromer} B.,  2021, \mn@doi [\aap]
  {10.1051/0004-6361/20214071110.48550/arXiv.2103.15112}, \href
  {https://ui.adsabs.harvard.edu/abs/2021A&A...649A.141U} {649, A141}

\bibitem[\protect\citeauthoryear{{Vaquero}, {Kovaltsov}, {Usoskin}, {Carrasco}
  \& {Gallego}}{{Vaquero} et~al.}{2015}]{vaquero_ea2015}
{Vaquero} J.~M.,  {Kovaltsov} G.~A.,  {Usoskin} I.~G.,  {Carrasco} V.~M.~S.,
  {Gallego} M.~C.,  2015, \mn@doi [\aap] {10.1051/0004-6361/201525962}, \href
  {https://ui.adsabs.harvard.edu/abs/2015A&A...577A..71V} {577, A71}

\bibitem[\protect\citeauthoryear{{Woolley}}{{Woolley}}{1963}]{Bradley}
{Woolley} Richard S.,  1963, \qjras, \href
  {https://ui.adsabs.harvard.edu/abs/1963QJRAS...4...47W} {4, 47}

\makeatother
\end{thebibliography}

\bsp	

\appendix

\section{Transit of Mercury}

\begin{figure}
\centering
\includegraphics[width=0.8\columnwidth]{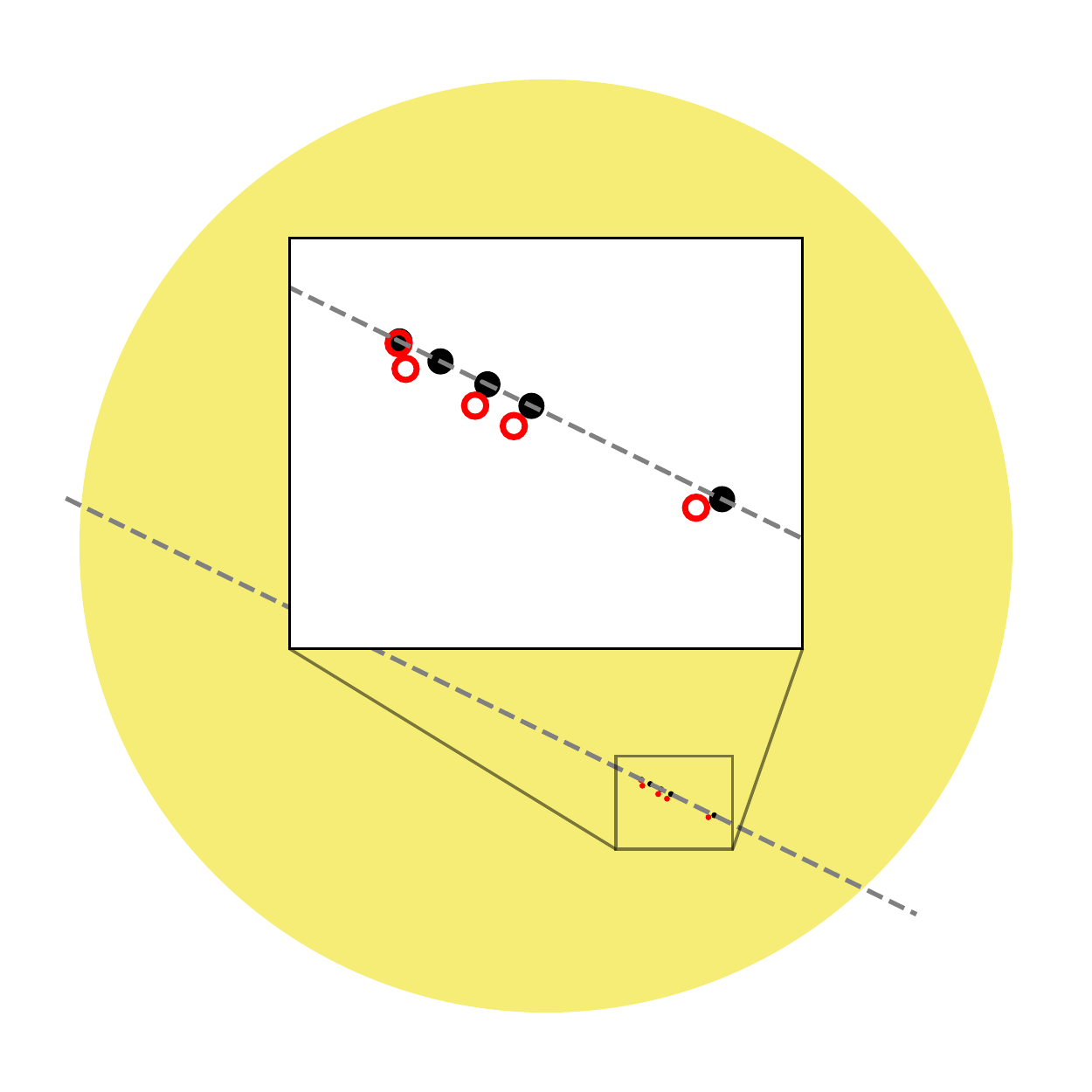}
    \caption{Transit of Mercury on 1799 May 7. The dashed line shows the path of Mercury derived with {\sc astropy}. The red circles show the positions of Mercury reconstructed from the observations. The black circles show the true position of Mercury, as a best fit to the historical records. The ratio of the coloured circles to the solar disk corresponds to the ratio of the apparent size of Mercury to the solar disk on that day (approximately 1/157). The white patch shows a magnified region of the Sun. The vertical points to the celestial north; Mercury moves from left to right with time.}
    \label{fig:merc}
\end{figure}

On 1799 May 7, Flaugergues observed the transit of Mercury across the solar disk with the cross-hair scheme (see sec.~\ref{sec:hv}). Using {\sc astropy}, we reconstructed Mercury's transit path and compared it to the positions of Mercury reconstructed from the observations. We used only the records that do not require additional information to obtain the coordinates, and skipped incomplete records. The five reconstructed positions are shown in red in Fig.~\ref{fig:merc}. The start time of the first reconstructed record is 14:35:30. We find that it best approximates the position of Mercury on 14:24:26~UTC. Thus we estimate the time difference between the Flaugergues's clock and UTC as 11~min~4~s. Using this estimate, we reconstructed Mercury's true position, consistent with subsequent records, and show them in black in figure~\ref{fig:merc}. 

Note that the planet disks are plotted to scale, with the ratio of the apparent size of Mercury to the solar disk on that day being approximately $1/157$. We conclude from Fig.~\ref{fig:merc} that errors in measuring the position of an object on the solar disk are of the order of the apparent size of Mercury which was about 12~arcsec in diameter.

In our opinion, several factors limit the accuracy of the measurements. The first one is the telescope's resolution, i.e.\ how blurred features on the Sun and the limb of the Sun appear to the observer. While there is some information on telescopes Flaugergues used, it is unclear which exactly he used for which observation. The second factor is the thickness of the wires forming the cross-hair scheme. Thirdly, the observation method is not instantaneous and requires the transit of the whole solar disk across each wire. It takes more than 3~minutes during which the relative position of the measured object may change. 

\label{lastpage}
\end{document}